\pdfoutput=1
\RequirePackage{ifpdf}
\ifpdf % We~are running pdfTeX in pdf mode
\documentclass[pdftex]{sigma}
\else
\documentclass{sigma}
\fi

\numberwithin{equation}{section}

\usepackage{pgf,tikz}
\usetikzlibrary{arrows,patterns,decorations.pathmorphing,backgrounds,positioning,fit,petri}
\usetikzlibrary {arrows.meta}

\newtheorem{Theorem}{Theorem}[section]
\newtheorem{Corollary}[Theorem]{Corollary}

\newtheorem{Proposition}[Theorem]{Proposition}
\newtheorem{Conjecture}[Theorem]{Conjecture}

{ \theoremstyle{definition}
\newtheorem{Definition}[Theorem]{Definition}

}

\usepackage{braket, xspace}
\DeclareMathOperator{\e}{e}
\DeclareMathOperator{\spn}{span}

\begin{document}
\allowdisplaybreaks

\newcommand{\arXivNumber}{2203.03633}

\renewcommand{\PaperNumber}{076}

\FirstPageHeading

\ShortArticleName{Tensors and Algebras: An Algebraic Spacetime Interpretation for Tensor Models}

\ArticleName{Tensors and Algebras: An Algebraic Spacetime\\ Interpretation for Tensor Models}

\Author{Dennis OBSTER}

\AuthorNameForHeading{D.~Obster}

\Address{Yukawa Institute for Theoretical Physics, Kyoto University,\\ Kitashirakawa, Sakyo-ku, Kyoto 606-8502, Japan}
\Email{\href{mailto:dennis.obster@yukawa.kyoto-u.ac.jp}{dennis.obster@yukawa.kyoto-u.ac.jp}}

\ArticleDates{Received April 24, 2022, in final form September 30, 2023; Published online October 18, 2023}

\Abstract{The quest for a consistent theory for quantum gravity is one of the most challenging problems in theoretical high-energy physics. An often-used approach is to describe the gravitational degrees of freedom by the metric tensor or related variables, and finding a way to quantise this. In the canonical tensor model, the gravitational degrees of freedom are encoded in a tensorial quantity $P_{abc}$, and this quantity is subsequently quantised. This makes the quantisation much more straightforward mathematically, but the interpretation of this tensor as a spacetime is less evident. In this work we take a first step towards fully understanding the relationship to spacetime. By considering $P_{abc}$ as the generator of an algebra of functions, we first describe how we can recover the topology and the measure of a compact Riemannian manifold. Using the tensor rank decomposition, we then generalise this principle in order to have a well-defined notion of the topology and geometry for a large class of tensors $P_{abc}$. We provide some examples of the emergence of a topology and measure of both exact and perturbed Riemannian manifolds, and of a purely algebraically-defined space called the semi-local circle.}


\Classification{83C45; 46C05; 16S15}

\section{Introduction}
One of the pinnacles of high-energy physics is the development of the standard model of particle physics. This model uses the framework of quantum field theory to describe the interaction of three of the four fundamental forces between elementary particles. One of the fundamental forces, gravity, has not yet successfully been added to this fundamental description of the universe. Most of the calculations of the standard model rely on perturbative renormalisation procedures, and this poses the biggest theoretical challenge in adding gravity to this picture. This is because general relativity, the theory that has been extremely successful in describing classical gravity~\cite{Abbott:2016blz,Dyson:1920cwa,EinsteinMercury1915,Will2014}, is perturbatively non-renormalisable, which makes the theory lose its predictive power at high energies~\cite{Goroff:1985th,tHooft:1974toh}. Experimentally it has been proven extremely difficult to actually do measurements in the quantum regime of gravity, since quantum effects are expected to play a role at the level of the Planck energy, a scale far ouch of reach for direct measurements currently.

One way to handle the perturbative renormalisation issues is to treat gravity in a non-perturbative way. The expectation of this approach is that, though the more straightforward perturbative renormalisation approaches fails to work for gravity, treating gravity in a non-perturbative way might solve these issues -- albeit mathematically more challenging. There are various approaches to this, all with some levels of success. For instance, one could try to reformulate renormalisation in a non-perturbative way and try to find a high-energy completion of the theory this way, which is the general strategy of the asymptotic safety programme~\cite{Reuter_PhysRevD.57.971, Reuter:2012id, Weinberg:1980gg}. Alternatively, one could consider a canonical quantisation approach as is in loop quantum gravity, reformulating general relativity using Ashtekar variables~\cite{rovelli1988new, Thiemann}. Another non-perturbative way of approaching quantum gravity is done by regularising the path integral using small building blocks, which is the fundamental idea behind for instance (causal) dynamical triangulation~\cite{Ambjorn:2012jv, Ambjorn:1998xu, Loll:2019rdj} and the usual tensor models~\cite{Ambjorn:1990ge, Godfrey:1990dt, Sasakura:1990fs}.

Tensor models are an interesting approach introduced as a generalisation of matrix models, which were successful in describing two-dimensional quantum gravity. The general idea is that, order-$d$ tensors generate $d$-dimensional space-times by gluing simplices together according to contractions between tensors, however it turns out that this does not result in the emergence of macroscopic spacetimes.\footnote{As an aside, there are also the so-called coloured tensor models which is argued to have a relationship to gravity, and furthermore studied for its connection to holography, due to the emergence of so-called melonic graphs~\cite{Bonzom:2011zz, Gurau:2009tw, Gurau:2013cbh}.} In the dynamical triangulation approach, causal dynamical triangulation seemed to fix many of these issues by introducing a notion of time by a causal requirement on the allowed triangulations. In tensor models, however, introducing such a restriction remains to be understood.\footnote{For some attempts to include a causal structure into tensor models, see~\cite{Eichhorn:2020sla, Jercher:2022mky}.} This led to the introduction of the canonical tensor model~\cite{Sasakura:2011sq,Sasakura:2013gxg}, which aims to describe a tensor model built from first principles in the Hamiltonian framework with algebraic similarities to the ADM-formalism of general relativity~\cite{Sasakura:2012fb}. This comes at the expense of a straightforward spacetime interpretation, though many connections to general relativity have been found implying that it can be interpreted as a model for quantum gravity~\cite{ Chen:2016ate, Sasakura:2014gia, Sasakura:2015pxa}. The mathematically straightforward way of quantising the model~\cite{Sasakura:2013wza}, and the interesting results from wave functions of the model~\cite{Kawano:2021vqc,Lionni:2019rty,Narain:2014cya,Obster:2017pdq,Obster:2017dhx,Obster:2020vfo, Sasakura:2013wza,Sasakura:2021lub, Sasakura:2019hql}, make it interesting to investigate.\looseness=-1

Contrary to some of the examples above, the canonical tensor model does not describe spacetime directly as a manifold with a (pseudo)-Riemannian metric or related variables. Instead, it uses a real $N$-dimensional tensor of order three as its fundamental variable, and one has to demonstrate the connection to gravity by the interpretation of this tensor and the dynamics of the model. Many of the difficulties when constructing a theory of quantum gravity can be traced back to the fact that this configuration space, for instance the configuration space of all $4$-dimensional metrics modulo the diffeomorphisms, is a very difficult configuration space to understand. The approach taken in this work is different, and the philosophy behind it partially overlaps with non-commutative geometry~\cite{vanSuijlekom2014noncommutative}, namely by describing a manifold through the algebra of functions on it. There has been some similar research done in this direction in the context of non-commutative geometry using spectral triples~\cite{Barrett:2019aig, Glaser:2019lck, Glaser:2019lcd}, though the setup differs slightly as the goal of this work is to find an interpretation for tensor models.

It is a well-known fact in algebraic geometry that compact Hausdorff spaces, $\mathcal{T}$, and the (real) algebra of functions on them, $C(\mathcal{T})$, are dual to each other through the Gelfand--Naimark theorem~\cite{gelfand1943imbedding}. Remarkably, for a smooth manifold $\mathcal{M}$ it is even possible to reconstruct the full smooth manifold structure, including charts and atlases, purely from the algebra of real smooth functions $C^\infty(\mathcal{M})$~\cite{nestruev2006smooth}. This means that knowing only the multiplication rules of elements of an \emph{abstract} algebra $\mathcal{A} \cong C^\infty(\mathcal{M})$ is enough to reconstruct the manifold. One important benefit is that, in the case of a compact Riemannian manifold, the algebra of smooth functions is a countably infinite-dimensional vector space, which as a configuration space is well understood. This is in contrast to for instance the configuration space of metrics over a compact Riemannian manifold, which is uncountably infinite-dimensional. There is one caveat here, namely that this only works for the topological degrees of freedom. In order to reconstruct the full (pseudo-)Riemannian manifold one needs more information, for instance through a spectral triple approach~\cite{connes1995, Connes:2008vs}.

This work introduces a framework connecting tensors to geometry in a different way than the original tensor models, namely by relating them to an algebra that is supposed to describe an algebra of functions, revisiting some of the ideas in~\cite{Sasakura:2005js, Sasakura:2006pq,Sasakura:2008pe, Sasakura:2007sv, Sasakura:2011qg, Sasakura:2011ma, Sasakura:2011nj}. The idea is that a tensor of order three, $P_{abc}$, describes an algebra as the structure coefficients of the algebra for basis-elements $\{f_a\}$
\begin{equation*}
 f_a \cdot f_b = \sum_c P_{abc}f_c.
\end{equation*}
An algebra that represents the pointwise product of functions ought to be commutative and associative. Guaranteeing commutativity is straightforwardly done by taking the tensor to be symmetric under permutations of the first two indices, since in that case
\begin{equation*}
 f_a \cdot f_b = \sum_c P_{abc} f_c = \sum_c P_{bac}f_c = f_b \cdot f_a.
\end{equation*}
However, guaranteeing associativity requires more work. If one wants an algebra corresponding to a pointwise product it is important to do this, because in a quantum theory for gravity using the tensor $P_{abc}$ as the fundamental variable, quantum perturbations might break the associativity. If, however, one would be able to link an associative algebra to the tensor $P_{abc}$, one would be able to link this algebra to a topological space according to methods introduced in~\cite{nestruev2006smooth}.

To this end, this paper defines the notion of the associative closure, in order to link a, either finite- or infinite-dimensional, tensor $P_{abc}$ to an associative algebra. This algebra might be infinite-dimensional, even if one starts with a finite-dimensional tensor. For example, if one starts with an algebra $\mathcal{F}$ which is $N$-dimensional, then the associative closure is realised by demanding that its $\mathbb{R}$-homomorphisms (i.e., the points of the space the functions are defined on) are consistent with all maximal sets of products of the algebra, later called partial algebras, that are associative already. Here, a maximal partial algebra means that there is no larger partial algebra that includes this set and is still associative. In practice this means that these maximal sets are treated as generators for the full algebra, i.e., the associative closure. For fully symmetric tensors $P_{abc}$ constructed from the algebra of square integrable functions on a Riemannian manifold, and which contain the structure coefficients of the full product of generating sets of this algebra, one can then reconstruct the algebra by generating tensor rank decompositions
\begin{equation*}
	P_{abc} = \sum_{i=1}^R \beta_i p_a^i p_b^i p_c^i,
\end{equation*}
such that $\beta_i>0$, and $R$ is minimal. The $\beta_i$ then correspond to the evaluation of the $L^2$-measure over a part of the topological space, and $p_a^i$ are points in the topological space. In the case of a Riemannian manifold, this means that part of the geometry of the Riemannian manifold, namely the measure, is recovered. Similarly to the spectral triple of non-commutative geometry, it will be argued that it is possible to include (and recover) the full geometric information of the (compact) Riemannian manifold.

This work is motivated by the canonical tensor model described above, as the original goal was to give a potential spacetime interpretation for it. As this model is set in the canonical (Hamiltonian) framework, the focus is on Riemannian manifolds since they are supposed to represent spatial slices of spacetime. It should be noted that it is not sure whether this is the most appropriate spacetime interpretation of the canonical tensor model, and this framework might be used for other models as well. Some potential implications to the canonical tensor model are discussed in this work, as there seem to be some interesting and encouraging consequences to using this interpretation. This framework also explains some of the topological and geometric results found before in the context of the canonical tensor model by using data analytic methods in~\cite{Kawano:2018pip}.

This work is organised as follows. In Section~\ref{sec:topology}, the duality between topological spaces and the algebra of functions on them is reviewed. It is then explained how one can construct a tensor from this algebra, and the example of the flat circle $S^1$ is introduced. If the reader is already aware with this duality, the section can be skipped with exception of Sections~\ref{sec:topology:example} and~\ref{sec:topology:CTM}, which are helpful for the general context of this work. Section~\ref{sec:associative} introduces the notion of an associative closure, to show how this framework can deal with tensors that do not correspond to associative algebras directly. Section~\ref{sec:associative_constr} then further develops this formalism, to show general ways in which an associative closure may be found. The example of the flat circle is revisited, and it is shown how from a five-dimensional symmetric tensor one can reconstruct the full algebra of smooth functions with a measure on it. Sections~\ref{sec:associative} and~\ref{sec:associative_constr} are mainly meant to introduce the mathematical foundations of this framework, and may be skipped on first read, though the examples in Section~\ref{sec:associative_constr} may still be useful to understand the framework better. In Section~\ref{sec:unit} is shown how one can generate a unit, if it is not trivially present in the algebra generated by the tensor yet. This is also where we identify a potential way to include more information about the geometry in a tensor. Section~\ref{sec:cases} then discusses a few examples of perturbations of the flat circle, an example of a purely algebraically-defined space called the semi-local circle, and the sphere. After this, in Section~\ref{sec:CTM} some implications to the canonical tensor model are discussed and finally Section~\ref{sec:summary} concludes this work.

\section[The duality between algebras and spaces, and the role of tensors]{The duality between algebras and spaces,\\ and the role of tensors}\label{sec:topology}

In this section, we briefly review the duality between smooth manifolds $\mathcal{M}$ and the real algebra of smooth functions $C^\infty(\mathcal{M})$, and develop an understanding of the role that tensors play in describing the algebra and the measure they induce. In the end of the section, we will describe an example of the circle. While we will mainly focus on compact Riemannian manifolds, much of this can be generalised to more general situations. In the following section, we introduce the definitions required to understand how these algebras can be constructed from finite-dimensional tensors. In Section~\ref{sec:associative_constr}, we apply these definitions and give explicit constructions to find the topological space and a measure corresponding to a finite-dimensional tensor.

Let us start by considering a smooth manifold $\mathcal{M}$. The real smooth functions on $\mathcal{M}$, denoted as $C^\infty(\mathcal{M})$, form an infinite-dimensional vector space, equipped with a pointwise product, together called the algebra of smooth functions on $\mathcal{M}$. In the following, we will show the duality between the space $\mathcal{M}$ and the algebra of smooth functions on it. In particular, we will reconstruct the set of points of the manifold and the topology. For further information on the reconstruction of the full smooth manifold structure (including charts and atlases) from the algebra of smooth functions we would like to refer to~\cite{nestruev2006smooth}.

Consider an abstract real unital associative commutative algebra $(\mathcal{F}, \cdot)$, where $\mathcal{F}$ denotes the linear space and $\cdot \colon \mathcal{F} \times \mathcal{F} \rightarrow \mathcal{F}$ the product operation. A linear map $p \colon \mathcal{F} \rightarrow \mathbb{R} $ is called an $\mathbb{R}$-algebra homomorphism if, besides the linearity conditions, it respects the product of the algebra, i.e., $\forall f,g \in \mathcal{F}, \, p(f \cdot g) = p(f) \cdot p(g)$, and maps the unit of $\mathcal{F}$ to the unit of $\mathbb{R}$. The dual space of the algebra is then defined as all $\mathbb{R}$-algebra homomorphisms of $\mathcal{F}$~\cite{nestruev2006smooth}:\footnote{$|\mathcal{F}|$ is often called the algebra of characters of $\mathcal{F}$.}
\begin{equation}\label{eq:duality:def_dual_space}
 |\mathcal{F}| := \{ p\colon \mathcal{F}\rightarrow \mathbb{R}\mid \forall f,g \in \mathcal{F},\, p(f\cdot g) = p(f) \cdot p(g) ;\, p(1_\mathcal{F}) = 1 \}.
\end{equation}
This dual space, henceforth called the algebraic dual space in order to differentiate it from the linear dual space, is the candidate ``set of points'', the main reason being that the evaluation maps $x \in \mathcal{M} , \, p_x(f) \equiv f(x)$ are naturally homomorphisms. An algebra is called geometric if the algebra $\mathcal{F}$ actually corresponds to functions on $|\mathcal{F}|$ by identifying $f(p) = p(f)$ (see \cite[Theorem, p.~24]{nestruev2006smooth}), and pointwise defined algebras are precisely that. Note that we will often use this identification in this work depending on the context, if we want to either view $p$ as a point or as a homomorphism.

\begin{figure}[t] \centering

\begin{tikzpicture}[smooth,scale=1]
%\node at (0,0.35) {\includegraphics[width=0.7\textwidth]{example.png}};
%\draw[very thin,color=gray,step=.5cm] (-6,-2.5) grid (6,3);
%\node[red] at (0,0) {$\circ$};
%\draw[color=black, thin, fill=gray] (-2.75,-0.6) ellipse (0.72 and 0.57);
%\draw[thick,gray] (-2,-0.4) -- (-3.4,-0.7);
\draw[color=black, thin,gray, fill=gray] (-2.05,-0.4) .. controls (-2.3,0.4) and (-3.6,0.1) .. (-3.4,-0.7) .. controls (-3.3,-1.3) and (-2.1,-1.5) .. (-2.05,-0.4);
\draw[thick,-Stealth,black] (-2.7,1.17) .. controls (-0.2,1.9)and (1.45,1.45) .. (2.5,-0.25);
\draw[thick,-Stealth,black] (3,-1.27) .. controls (1.5,-1.85)and (-1,-1.5) .. (-2.5,-0.71);
\draw[thin,black] (-5.35,-1.45) .. controls (-4.56,-0.2) and (-4.13,1.3) .. (-3.93,2.5);
\draw[thin,black] (-2.6,-2.2) .. controls (-1.6,-0.95) and (-1.2,0.55) .. (-1.2,1.74);
\draw[thin,black] (-5.35,-1.45) .. controls (-4.56,-1.23) and (-3.5,-1.4) .. (-2.6,-2.2);
\draw[thin,black] (-3.93,2.5) .. controls (-3,2.44) and (-1.5,2.25) .. (-1.2,1.74);
\draw[thick,black] (0.85,-1) -- (5.4,-1);
\draw[thick,black] (2.75,-0.8) .. controls (2.69,-1) .. (2.75,-1.2);
\draw[thick,black] (3.55,-0.8) .. controls (3.61,-1) .. (3.55,-1.2);
\node[black] at (1.2,1.5) {$f\in\mathcal{F}$};
\node[black] at (5.2,-0.75) {$\mathbb{R}$};
\node[black] at (3.15,-1.25) {$V$};
\node[black] at (-3.6,2.18) {$|\mathcal{F}|$};
\node[black] at (-1,-1.7) {$f^{-1}(V)$};
\end{tikzpicture}

 \caption{An illustration of the construction of the topology on $|\mathcal{F}|$.}
 \label{fig:duality:topology}
\end{figure}
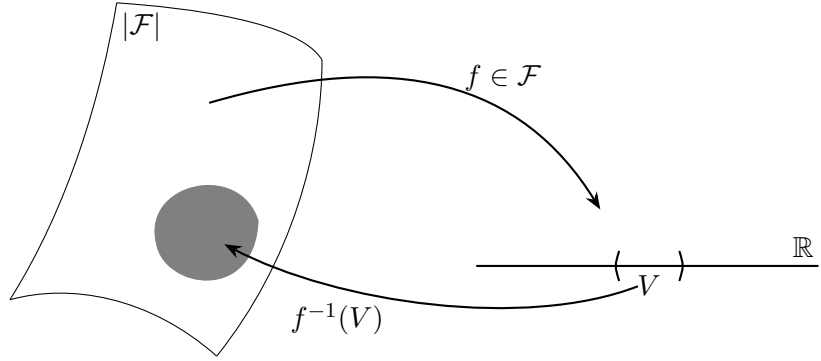
The topology on $|\mathcal{F}|$ is now defined as the weakest topology such that the functions $f\in\mathcal{F}$ become continuous, inheriting the continuity from $\mathbb{R}$. To make this precise, we can construct a basis for the topology on $|\mathcal{F}|$ by taking all open subsets $V\subset\mathbb{R}$, and using $f^{-1}(V)$ for all $f\in\mathcal{F}$ as a basis for the topology. For an illustration of this, see Figure~\ref{fig:duality:topology}. Taking the algebra of smooth functions $C^\infty(\mathcal{M})$, we can thus reconstruct $\mathcal{M} = |C^\infty(\mathcal{M})|$.

As mentioned before, it is possible to reconstruct the full topological manifold structure from the algebra of smooth functions. In this work, we will not go deeper into this, as we are mainly interested here in the reconstruction of the topology and measure from a tensor. It is worth noting however that we are mainly interested in algebras that correspond to smooth algebras of compact Riemannian manifolds. This has profound implications, as a compact Riemannian manifold carries a natural measure, locally infinitesimally ${\rm d}^d x \, \sqrt{\det q(x)}$ where $q$ denotes the metric tensor field, $d$ the dimension of the Riemannian manifold, and $x$ the local coordinates. Because of this we can define a natural inner product on the space of smooth functions, called the $L^2$-inner product:
\begin{equation}\label{eq:topology:innerp}
 \braket{f | g} := \int_\mathcal{M} {\rm d}^d x \, \sqrt{\det q(x)} f(x) g(x).
\end{equation}
On a compact Riemannian manifold, we arrive at the square integrable functions $L^2(\mathcal{M})$ by taking the closure of $C^\infty(\mathcal{M})$ with respect to the inner product above. $L^2(\mathcal{M})$ is a countably infinite dimensional Hilbert space, which means that we can choose an orthonormal Schauder basis
\begin{equation*}%\label{eq:duality:basis}
 \big\{ f_a \, \in \, L^2(\mathcal{M}) \mid a\geq 1 \big\}.
\end{equation*}
This will be important below. Since one can always extend the algebra to the full Hilbert space, we are assuming that the linear space is a Hilbert space below.

One may now define a tensorial quantity using this basis and the algebra product
\begin{equation}\label{eq:def_P:innerp}
 P_{ab}{}^c := \braket{f_c | f_a \cdot f_b }.
\end{equation}
Another way to view this tensor is as the structure coefficients of the algebra, as
\begin{equation}\label{eq:def_P:algebra}
 f_a \cdot f_b =: \sum_{c\geq 1} P_{ab}{}^c f_c.
\end{equation}
Note that using this decomposition, one can recover every possible product in the algebra. Taking $f,g\in \mathcal{F}$, with $f=\sum_{a\geq 1} \alpha^a f_a$ and $g=\sum_{b\geq 1} \beta^b f_b$ gives
\begin{equation*}
 f\cdot g = \sum_{a\geq 1}\sum_{b\geq 1} \alpha^a \beta^b f_a \cdot f_b = \sum_{a,b,c \geq 1} \alpha^a \beta^b P_{ab}{}^c f_c.
\end{equation*}
Equation~\eqref{eq:def_P:algebra} lends itself to find the dual space~\eqref{eq:duality:def_dual_space}, as the elements of this dual space are a subset of the linear dual space $\mathcal{F}^*$.\footnote{Note that here we call the linear dual space the space of all (bounded) linear functionals, while the algebraic dual space are the homomorphisms of the algebra as defined in~\eqref{eq:duality:def_dual_space}.} To do this, consider a homomorphism $p \colon \mathcal{F} \rightarrow \mathbb{R}$. The linear dual space is isomorphic to $\mathcal{F}$, as we are assuming a Hilbert space structure, so we can use the dual elements of the basis $\{ f_a \}$, let us denote these elements by $\{\alpha^a\}$ such that $\alpha^a(f_b) = \delta^a_b$, as a basis of $\mathcal{F}^*$. Specifically we can write $p$ as $p = \sum_{a\geq 1} p_a \alpha^a$. In terms of our original basis, the components $p_a$ can now be found by evaluating $p(f_a) = \sum_{b\geq 1} p_b\, \alpha^b(f_a) = p_a $. Given $P_{ab}{}^c$, we can find the algebraic dual space of the algebra by restricting $p\in \mathcal{F}^*$ to the hypersurface where
\begin{equation}\label{eq:duality:find_homo}
 p_a p_b = \sum_{c \geq 1} P_{ab}{}^c p_c,
\end{equation}
and we now understand that these components can be understood as evaluation maps of the basis functions $f_a(p) \equiv p(f_a) = p_a$.

The idea is to build a theory around these coefficients $P_{ab}{}^c$ in the form of a tensor, where changes in the coefficients change the algebra and thus affect the corresponding manifold. A~thing to note is that, since the algebra is commutative and the product operation on the pointwise defined algebras discussed here is self-adjoint, the structure coefficients will be totally symmetric: $P_{ab}{}^c = P_{ca}{}^b = P_{bc}{}^a = P_{ba}{}^c = P_{cb}{}^a = P_{ac}{}^b$. This means that constructing a theory that guarantees the emergence of commutative algebras with an inner-product structure is equivalent to focusing on totally symmetric tensors, as is done in the canonical tensor model for instance.

Until now we have been careful to always write $P_{ab}{}^c$ with one upper index and two lower, but since we have a Hilbert space structure with a totally symmetric tensor it is not really necessary to keep this upper index most of the time, and simply write $P_{abc}$. One can view this as ``lowering the index'' with a metric $g_{ab} = \braket{f_a | f_b} = \delta_{ab}$ as
\begin{equation*}
 P_{abc} = g_{cd} \, P_{ab}{}^d.
\end{equation*}
In this work, we will usually use the lower-index notation $P_{abc}$ as this is also the notation used in the canonical tensor model. However, in some cases we will still write $P_{ab}{}^c$ if we wish to emphasise the relationship to the linear dual space $\mathcal{F}^*$.

For the correspondence between tensors and spaces mentioned above to lead to a theory of gravity there is some more work to be done. The main factors we will address here are as follows:%
\begin{itemize}\itemsep=0pt
 \item In practice, one can never do practical calculations (e.g., on a computer) for infinite-dimensional tensors. Instead, one expects that defining a model for finite-dimensional $N$, and either considering this to be a fundamental feature of the universe or taking some $N\rightarrow\infty$ limit. Either way, it will be useful to have some procedure to connect a finite-dimensional non-associative algebra to an associative infinite-dimensional algebra. In this work we try to resolve this by looking for an associative closure of the algebra, such that a tensor $P_{ab}{}^c$ actually corresponds to an associative algebra in Section~\ref{sec:associative}.
 \item Quantum perturbations might affect the algebra such that it is not associative anymore. However, algebras of pointwise defined functions are inherently associative, hence there is a need to treat these algebras if one wants to stick to this interpretation. The associative closure mentioned above will also be able to treat these algebras.\footnote{Note that this could lead to two different interpretations for a model, one where the non-associative perturbations actually lead to a different associative algebra, and one where the non-associativity is seen as fundamental.}
 \item A theory for gravity should actually influence the metric on the Riemannian manifold, not just the topology. We take a first step to this in Section~\ref{sec:associative_constr}, where we reproduce the measure on the manifold.
 \item Similarly, not every algebra has a well-defined unit. We will describe how to generate a~unit in an algebra in Section~\ref{sec:unit}, and we will also see that this gives us an opportunity to describe more (or all) of the geometric information of the Riemannian manifold.
\end{itemize}

\subsection{Example: The flat circle}\label{sec:topology:example}
In this subsection, we introduce the example of the flat circle. The flat circle is the example we will use to develop our understanding of the representation of tensors, after which we will consider different algebras later on in Section~\ref{sec:cases}. The flat circle is given by the 1-dimensional circle $S^1$ with a flat metric on it. One choice of basis for the smooth functions on the 1-dimensional circle is given by functions of the form\footnote{These are exactly the eigenfunctions of the Laplace--Beltrami operator on the circle, which will be discussed in Section~\ref{sec:unit}.} (normalised according to the $L^2\big(S^1\big)$ inner product)
\begin{gather}\label{eq:topology:example:basis}
 \left\{f_1 \!=\! \frac{1}{\sqrt{2\pi}}, \, f_2 \!= \!\frac{1}{\sqrt{\pi}}\sin(x),\, f_3 \!= \!\frac{1}{\sqrt{\pi}}\cos(x) ,\, f_4 \!=\! \frac{1}{\sqrt{\pi}}\sin(2 x) ,\, f_5 \!=\! \frac{1}{\sqrt{\pi}}\cos(2 x), \dots \right\},\!\!
\end{gather}
with $x\in(0,2\pi]$. By evaluating the product of the basis-functions above, we can find the structure constants by using~\eqref{eq:def_P:algebra}. For example, from the product
\begin{equation}\label{eq:duality:example_product}
 f_2 \cdot f_2 = \frac{1}{\pi} \sin(x) \cdot \sin(x) = \frac{1}{2 \pi}\left(1-\cos(2 x)\right) = \frac{1}{\sqrt{2\pi}} f_1 - \frac{1}{2\sqrt{\pi}} f_5,
\end{equation}
we find
\begin{equation*}
 P_{22}{}^1 = \frac{1}{\sqrt{2\pi}}, \qquad P_{22}{}^5 = - \frac{1}{2\sqrt{\pi}} , \qquad P_{22}{}^a =0 \qquad \text{for} \ a \neq 1,5.
\end{equation*}
Actually, we can find the entries of the tensor by considering the basic multiplication rules
\begin{gather}
 \sin(n\, x) \cdot \sin(m\, x) = \frac{1}{2} (\cos ((m-n)x ) - \cos ((m+n) x ) ),\nonumber\\
 \cos(n\, x) \cdot \sin(m\, x) = \frac{1}{2} (\sin ((m-n)x ) + \sin ((m+n) x ) ),\nonumber\\
 \cos(n\, x) \cdot \cos(m\, x) = \frac{1}{2} (\cos ((m-n)x ) + \cos ((m+n) x ) ).\label{eq:duality:all_products}
\end{gather}
Translating this to the basis elements one finds the multiplication laws for $a>b$,
\begin{gather*}%\label{eq:duality:all_products_basis}
f_{2a}f_{2b} = \frac{1}{2\sqrt{\pi}}( f_{2(a-b)+1} - f_{2(a+b)+1}),\\
f_{2a}f_{2b+1} = \frac{1}{2\sqrt{\pi}} ( f_{2(a+b)} - f_{2(a-b)} ) , \\
f_{2a+1}f_{2b+1} = \frac{1}{2\sqrt{\pi}} ( f_{2(a-b)+1} + f_{2(a+b)+1}).
\end{gather*}
From this we can write out the full tensor $P_{ab}{}^c$. Alternatively we can find these components by evaluating~\eqref{eq:def_P:innerp}, for instance,
\begin{equation*}
 P_{22}{}^5 = \frac{1}{\pi^{3/2}}\int_{0}^{2\pi} {\rm d}x\, \sin(x) \sin(x) \cos(2x) = -\frac{1}{2\sqrt{\pi}}.
\end{equation*}
The claim of this section is now that we can reconstruct the topology of the circle, just by knowing the tensor $P_{ab}{}^c$. To further convince ourselves of this, let us try to find all the homomorphisms of the first few functions $\{1, \sin(x), \cos(x) \}$. By inspecting~\eqref{eq:duality:all_products}, we can see that all information of the products of these functions is already present if we restrict the tensor $P_{ab}{}^c$ to labels up to $a,b,c \leq 5$. This restriction effectively means the tensor acts on $\left.\mathcal{F}=C^\infty\big(S^1\big)\right|_{5}$ which is then isomorphic to $\mathbb{R}^5$. We call the dimension of the real vector space that such a tensor acts on $N$. In this work, if we say $N$-dimensional tensor of order $d$, we mean a tensor acting on $d$ copies of a $N$-dimensional real vector space. If not mentioned, the order of the tensor is meant to me $d=3$. In Section~\ref{sec:associative}, we will properly define some of the notions described below by considering partial algebras.

\begin{figure}[t]
 \centering
 \begin{minipage}{0.3\textwidth}
 \includegraphics[width=0.95\textwidth]{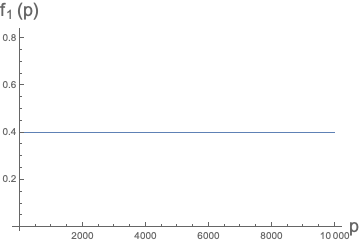}
 \end{minipage}
 \begin{minipage}{0.3\textwidth}
 \includegraphics[width=0.95\textwidth]{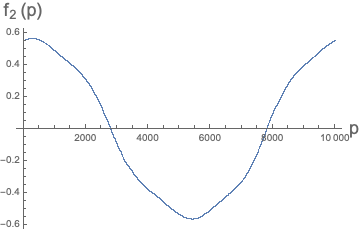}
 \end{minipage}
 \begin{minipage}{0.3\textwidth}
 \includegraphics[width=0.95\textwidth]{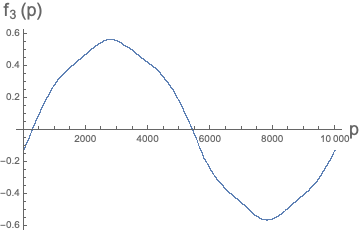}
 \end{minipage}

 \caption{The homomorphisms of the first three functions $f_1$, $f_2$, $f_3$ of the circle. As described in the text, we calculated $10000$ homomorphisms for these functions, and plotted the list $p^i(f_a)$. The homomorphisms where calculated using \textsc{Mathematica} by minimising $\big(p_ap_b - \sum_{c=1}^5 P_{ab}{}^c p_c\big)^2$ and only keeping the ones that evaluated to zero (excluding the trivial zero-map), then they were ordered as described in~\eqref{eq:duality:example:topology_norm} (note that the first point is arbitrary, thus there is an arbitrary shift). This is supposed to represent a topological circle, so slight deformations are merely an artifact of the random initial conditions procedure for the minimizing function of \textsc{Mathematica}. The rigid circle will be found once we take the measure into account in Section~\ref{sec:associative_constr}.}
 \label{fig:duality:circle_topology}
\end{figure}

Finding the set of points of $S^1$ can then be done by finding all solutions to~\eqref{eq:duality:find_homo}. Specifically, as we only consider the first three functions here,
\begin{equation}\label{eq:duality:example_homo}
 p\in \mathcal{F}^* \cong \mathbb{R}^5,\, a,b \in \{1,2,3\} \colon \ p_a p_b = \sum_{c=1}^5 P_{ab}{}^c p_c.
\end{equation}
All $p$ satisfying the above equation yield $|\mathcal{F}|$ as in~\eqref{eq:duality:def_dual_space}. Then, in order to introduce a topology on $|\mathcal{F}|$ we use the construction explained below~\eqref{eq:duality:def_dual_space}. To illustrate this in a figure, we calculated $10000$ points that satisfy~\eqref{eq:duality:example_homo}, which are labeled by $p^i$, $i\in\{1, \dots, 10000\}$. In order to visualise the topology, consider the elementary definition of continuity,
\begin{equation*}
 \lim_{\epsilon\rightarrow 0^+} f_a(p) - f_a(p+\epsilon) \rightarrow 0.
\end{equation*}
Of course, when only considering a finite number of points, this will never go to zero exactly. However, it does make sense to visualise this by arranging the points such that the absolute value of
\begin{equation*}
 f_a\big(p^i\big) - f_a\big(p^{i+1}\big) = p_a^i - p_a^{i+1}
\end{equation*}
is as small as possible. To do this, start with a point, say $p^1$. Then, take $p^2$ to be the point that minimises $\sum_{a=1}^5 \big|p_a^1-p_a^2\big|^2$. This is then done for all points going forward, such that
\begin{equation}\label{eq:duality:example:topology_norm}
 \sum_{a=1}^5 \big|p_a^i-p_a^{i+1}\big|^2
\end{equation}
is minimised for every $i < 10000$. The result of this is shown in Figure~\ref{fig:duality:circle_topology}, note that the form of these functions really resembles the functions $\big\{\frac{1}{\sqrt{2\pi}}, \frac{1}{\sqrt{\pi}}\sin(x), \frac{1}{\sqrt{\pi}} \cos(x)\big\}$ up to deformations and a phase.

\subsection{Tensor rank decompositions and persistent homology}\label{sec:topology:CTM}
This short subsection describes the appearance of some similarities to the work in~\cite{Kawano:2018pip}, which has seen the furthest development so-far of a geometric interpretation of the canonical tensor model. These similarities were the main motivation of the present work, as the work has been useful in further analysis~\cite{Sasakura:2019hql} and this work gives a possibility to understand the appearance of these elements to a more fundamental level.

In~\cite{Kawano:2018pip}, no reference was made to algebras of functions. Instead, the set of points, $|\mathcal{F}|$, of the (topological) space corresponding to the tensor defined by the elements of the tensor rank decomposition
\begin{equation*}
	P_{abc} = \sum_i^R p_a^i p_b^i p_c^i,
\end{equation*}
where $q_a^i \in \mathbb{R}^N$. For smooth spaces, it was proposed that this means that there are many such tensor rank decompositions possible. In Section~\ref{sec:associative_constr}, the appearance of the tensor rank decomposition is explained from the space of functions over a Riemannian manifold.

After defining the set of points, the persistent homology technique was used in order to heuristically couple topologies to these tensors~\cite{Kawano:2018pip} as follows. First, an inner product on the set of points was introduced for $p^i, p^j \in |\mathcal{F}|$ as
\begin{equation*}
 p^i \cdot p^j = \sum_{1}^N p_a^i p_a^j.
\end{equation*}
Subsequently, points were argued to be ``close'' if their inner product was large. A graph was built, connecting elements $i$ and $j$ if the inner product was above a certain threshold. The graph distance was then used as the distance function for persistent homology.

The intuitive statement that points are close if the inner product between two points is large, makes total sense from the point of view of the present work to understand the topological properties of the space: it is nothing but minimising~\eqref{eq:duality:example:topology_norm}. This suggests that the approach used in~\cite{Kawano:2018pip} was indeed valid to find the topological properties of a space.

\section{Defining the associative closure}\label{sec:associative}
In the previous section we developed an understanding for the emergence of a topology from infinite-dimensional tensors, and in the example of the flat circle we also already got a glimpse of the emergence of a topology from just a finite subset of such an algebra. In this section, we will develop this further, where we will define the associative closure of a (either finite- or infinite-dimensional) tensor $P_{abc}$. After this, in Section~\ref{sec:associative_constr}, we will describe how to construct an associative closure for finite-dimensional tensors.

First, let us explain why we want to look for an associative closure of a tensor $P_{abc}$. Consider an $N$-dimensional tensor $P_{abc}$, and we assume this tensor generates an algebra
\begin{equation*}%\label{eq:associative:finite_tensor_algebra}
 f_a \cdot f_b = \sum_{c=1}^N P_{abc}f_c.
\end{equation*}
If this tensor $P_{abc}$ is defined as in Section~\ref{sec:topology}, where $N$ acts like a cutoff value, this will naturally be a non-associative algebra.

Let us explicitly show the non-associativity in the case of the five-dimensional algebra of the flat circle of Figure~\ref{fig:duality:circle_topology}. One example of a non-associative product is
\begin{gather*}
 f_2 \cdot (f_3 \cdot f_4) = f_2 \cdot \sum_{a=1}^5 P_{34a}f_a = \frac{1}{2\sqrt{\pi}} f_2\cdot f_2 = \frac{1}{4\pi} \big(\sqrt{2}f_1 - f_5\big),\\
 (f_2 \cdot f_3) \cdot f_4 = \sum_{a=1}^5 P_{23a}f_a \cdot f_4 = \frac{1}{2\sqrt{\pi}}f_4 \cdot f_4 = \frac{1}{2\sqrt{2}\pi}f_1.
\end{gather*}
It can be seen that this non-associativity is due to the cutoff introduced by a finite $N$, as for $C^\infty\big(S^1\big)$
\begin{gather*}
 \sin(x)(\cos(x) \sin(2x)) = \frac{1}{2} \sin(x) (\sin(x)+ \sin(3x)) = \frac{1}{4}(1 - \cos(4x)),\\
 (\sin(x) \cos(x)) \sin(2x) = \frac{1}{2} \sin(2x) \sin(2x) = \frac{1}{4} (1-\cos(4x)).
\end{gather*}
In the finite-dimensional case the $\sin(3x)$ contribution ``drops out'', because of which the product is not associative anymore.

This shows how, even if one does not intend to describe any non-associative effects, non-associativity shows up in this description. In practice, one will not be able to do any calculations with infinite $N$, so a way to connect a finite-dimensional tensor $P_{abc}$ to an infinite-dimensional associative algebra is necessary in order to know which geometric space one is talking about.

To properly define the associative closure, we first need to introduce some other notions first.
\begin{Definition}
 Consider an algebra, i.e., a vector space $\mathcal{F}$ with a bilinear product operator $\mathcal{P} \colon \mathcal{F}\times \mathcal{F} \rightarrow \mathcal{F}$. A sub-vector space $\mathcal{S}\subset\mathcal{F}$ with the restricted bilinear product operator $\mathcal{P} \colon \mathcal{S}\times \mathcal{S} \rightarrow \mathcal{F}$ is called a \emph{partial algebra} of $\mathcal{F}$. The partial algebra is unital if $\exists\, 1\in\mathcal{S}$, $\forall f\in\mathcal{S} \colon \mathcal{P}(1,f) = f$. Similarly, the partial algebra is commutative if $\forall f,g\in\mathcal{S}$, $\mathcal{P}(f,g) = \mathcal{P}(g,f) $ and associative if $\forall f,g,h\in\mathcal{S}$, $\mathcal{P}(f,\mathcal{P}(g,h)) = \mathcal{P}(\mathcal{P}(f,g),h)$.
\end{Definition}

Note that for the partial algebra, for the commutativity and associativity conditions, the result of $\mathcal{P}(f,g)$ (and the others) can be lie outside of $\mathcal{S}$, in $\mathcal{F}$. For a finite-dimensional or countably infinite dimensional partial algebra, the operator $\mathcal{P}$ may be decomposed as a tensor~$P_{abc}$ in a~similar fashion as described in Section~\ref{sec:topology}.

\begin{Definition}\label{def:associative:dual_space}
 The \emph{algebraic dual-space} of a \emph{partial} algebra $\mathcal{S}\subset\mathcal{F}$ with operator $\mathcal{P}$, denoted~$|\mathcal{S}|$, are all $p\in\mathcal{F}^*$ that are homomorphisms of the partial algebra in the sense that
 \begin{equation*}
 |\mathcal{S}| := \{ p \in \mathcal{F}^* \mid \forall f,g \in \mathcal{S},\, p\left(\mathcal{P}(f,g)\right) = p(f) p(g); \, p(1_\mathcal{F}) = 1 \}.
 \end{equation*}

\end{Definition}
Note that while the definition is similar to~\eqref{eq:duality:def_dual_space}, the crucial differences here are that the actual product of two vectors may lie outside of the partial algebra, and the elements $p\in|\mathcal{S}|$ are elements of $\mathcal{F}^*$, but not $\mathcal{S}^*$, in general. In a sense, we are looking at sub-vector spaces with a product that is closed not in itself, but in the whole vector space. The general idea is that while we want to find an associative closure, we will not want to alter the part of the algebra that is already associative.

Let us go back to the example of the flat circle introduced in Section~\ref{sec:topology}. Here we discussed the tensor $P_{abc}$ acting on a five-dimensional vector space. In the language introduced here, the algebra discussed is given by $\mathcal{F}\cong \mathbb{R}^5$, and there is a partial algebra $\mathcal{S} = \text{span} \{f_1, f_2, f_3\}\cong \mathbb{R}^3$. This partial algebra can be verified to be unital, commutative and associative. Its algebraic dual space, $|\mathcal{S}|\subset\mathbb{R}^5$, is discussed in Figure~\ref{fig:duality:circle_topology}. The original algebra $C^\infty\big(S^1\big)$ would correspond to the associative closure we wish to define below.

In order to be able to describe more general systems than those with an obvious candidate for a partial algebra as the flat circle, we need some more notions. They will be introduced here, but later on in Section~\ref{sec:cases} their use will become more clear.

\begin{Definition}
 A \emph{system} of partial algebras is a set of partial algebras $\{\mathcal{S}_i \mid i\in \mathcal{I}\}$, where $\mathcal{I}$ is some index set, such that for every pair $(\mathcal{S}_i,\mathcal{S}_j)$ their (algebraic) dual spaces have a nontrivial intersection, i.e.,
 \begin{equation*}
 |S_i| \cap |S_j| \neq \varnothing.
 \end{equation*}
 The \emph{algebraic dual-space} of a system of partial algebras $\{\mathcal{S}_i\}$ is the intersection of all of the dual spaces, i.e.,
\begin{equation*}
 |\{\mathcal{S}_i\}| := \bigcap_i |\mathcal{S}_i|.
\end{equation*}
\end{Definition}

\begin{Definition}
 The \emph{range} of a partial algebra $\mathcal{S}$ is defined as the sub-vector space $\mathcal{K}^{(
 \mathcal{S})}$ of $\mathcal{F}$, $\mathcal{S}\subset\mathcal{K}^{(\mathcal{S})}\subset\mathcal{F}$, which is reached by evaluating products of elements of $\mathcal{S}$:
\begin{equation*}
 \mathcal{K}^{(\mathcal{S})} := \{\mathcal{P}(f,g) \mid f,g \in \mathcal{S}\}.
\end{equation*}
The \emph{range of a system of partial algebras} is defined as the union of the ranges of the partial algebras
\begin{equation*}
 \mathcal{K}^{(\{\mathcal{S}_i\})} := \bigcup_{\{i\in\mathcal{I}\}} \mathcal{K}^{(\mathcal{S}_i)}.
\end{equation*}
Note that this is the union and not the sum of vector spaces.
\end{Definition}

\begin{Definition}
 A system of unital commutative associative partial algebras $\{\mathcal{S}_i\}$ of $\mathcal{F}$ is called \emph{maximal} in the set of all unital commutative associative partial algebras of $\mathcal{F}$ if there is no system of unital commutative associative partial algebras of $\mathcal{F}$, say $\{\mathcal{T}_j\}$, with a larger range, i.e., $ \mathcal{K}^{(\{\mathcal{S}_i\})} \subset \mathcal{K}^{(\{\mathcal{T}_j\})} \Rightarrow \mathcal{K}^{(\{\mathcal{S}_i\})} = \mathcal{K}^{(\{\mathcal{T}_j\})}$. A system is said to be \emph{covering} $\mathcal{F}$ if its range spans the whole algebra, i.e., $\mathcal{K}^{(\{\mathcal{S}_i\})} = \mathcal{F}$.
\end{Definition}

In the example of the flat circle, we have several unital commutative associative partial algebras. Some examples are $\mathcal{S}_0 = \spn\{1\}$, $\mathcal{S}_1 = \spn\{1, \sin(x)\}$, $\mathcal{S}_2 = \spn\{1, \cos(x)\}$, $\mathcal{S}_3 = \spn\{1,\sin(x),\cos(x)\}$. Their ranges are $\mathcal{K}_0=\spn\{1\}$, $\mathcal{K}_1=\spn\{ 1, \sin(x), \cos(2x)\}$, $\mathcal{K}_2=\spn\{1, \cos(x), \cos(2x)\}$ and $\mathcal{K}_3=\spn\{1, \sin(x), \cos(x), \sin(2x), \cos(2x)\}$, respectively. An example of a non-trivial system would be the system of $\mathcal{S}_1$ and $\mathcal{S}_2$ though they are not maximal nor covering. However, the system of only $\mathcal{S}_3$ is a covering system of $\mathbb{R}^5$.

Another equivalent covering system would be all $\mathcal{S}_\alpha = \{1, f_\alpha \equiv \sin(\alpha) \sin(x) + \cos(\alpha) \cos(x)\}$ for $\alpha \in [0, 2\pi)$. Note that every partial algebra in this system is two-dimensional, but taking all of them together will still cover all of $\mathcal{F}$ since $f_\alpha \cdot f_\alpha = (\sin(\alpha)\sin(x) + \cos(\alpha) \cos(x))(\sin(\alpha)\sin(x) + \cos(\alpha) \cos(x)) = \frac{\sin(\alpha)^2}{2}(1 - \cos(2x)) + \sin(\alpha)\cos(\alpha) \sin(2x) + \frac{\cos(\alpha)^2}{2}(1 + \cos(2x))$. Therefore, the range of $\mathcal{S}_\alpha$ is
\begin{equation*}
 \mathcal{K}^{(\mathcal{S}_\alpha)} = \spn \{ 1, \sin(\alpha) \sin(x) + \cos(\alpha) \cos(x), \sin(2\alpha)\sin(2x) + \cos(2\alpha) \cos(2x) \},
\end{equation*}
yielding a range for the system of
\begin{equation*}
 \mathcal{K}^{(\{\mathcal{S}_\alpha\})} = \bigcup_{\alpha} \mathcal{K}^{(\mathcal{S}_\alpha)} = \mathcal{F}.
\end{equation*}

The reason why we are introducing all this terminology with respect to partial algebras is twofold. Firstly, we wish to properly define a way to find potential ``signs of associativity'', since a pointwise algebra as discussed in Section~\ref{sec:topology} should be associative. These should be a~good starting point of constructing an associative closure. Secondly, we would like these signs of associativity to actually still be present in the full associative closure. If we for instance consider the covering partial algebra in Figure~\ref{fig:duality:circle_topology}, we would like these functions to look the same in the final associative closure and not look wildly different.

We now turn to the more general situation where we have a (finite-dimensional or countably infinite-dimensional) Hilbert space $\mathcal{F}$ with a basis $\mathcal{B}_\mathcal{F} = \{ f_a \mid a\geq 1\}$ and a symmetric tensor $P_{abc}$ acting on it. This tensor will certainly span an algebra by the rule~\eqref{eq:def_P:algebra}, however this algebra might not be unital nor associative. Here we will assume that this algebra spanned is already unital, i.e., it contains a unit $\mathcal{F} \ni 1 \equiv \alpha_a f_a$ \big(using the Einstein summation convention\footnote{In this work, whenever repeated indices are used the Einstein-summation convention should be assumed, unless specified otherwise.}\big) such that
\begin{equation*}
 \forall f_b\in\mathcal{B}_\mathcal{F} ,\ 1 \cdot f_b = \alpha_a P_{abc} f_c = f_b \
 \Rightarrow \ \alpha_a P_{abc} = \delta_{bc}.
\end{equation*}
A tensor $P_{abc}$ that generates a unital algebra as above is called a \emph{unital} tensor. In Section~\ref{sec:unit}, we will describe a procedure to construct such a unit in many cases, but for now we assume that the tensor $P_{abc}$ already contains it. Note that the unit in the example of the tensor representing the flat circle is given by $\mathbf{1} = \sqrt{2\pi} f_1$.

We now define an associative extension of a tensor $P_{abc}$ as follows
\begin{Definition}\label{def:associative:ass_extension}
 An associative extension of a tensor $P_{abc}$ acting on a countable Hilbert space~$\mathcal{F}$ with basis $\mathcal{B}_\mathcal{F}$ is an algebra $(\mathcal{A},\cdot)$, consisting of a countable Hilbert space~$\mathcal{A}$, which is an extension of $\mathcal{F}\subset\mathcal{A}$, and a product operation $\cdot\colon \mathcal{A} \times \mathcal{A} \rightarrow \mathcal{A}$ satisfying:
 \begin{enumerate}\itemsep=0pt
 \item The algebra is unital, associative and commutative.
 \item The product operation $\cdot$ reduces to $P_{abc}$ on $\mathcal{F}$ in the sense that
 \begin{equation*}
 \forall f_a,f_b,f_c \in \mathcal{B}_\mathcal{F} \subset \mathcal{A} \colon \ P_{abc} = \braket{f_c | f_a \cdot f_b}.
 \end{equation*}
 \item Every element of the algebraic dual space $p\in|\mathcal{A}|$ projected to $\mathcal{F}^*$, i.e., $p|_{\mathcal{F}^*}$, lies in the dual space of some maximal system of partial algebras $|\{\mathcal{S}_i\}|\subset\mathcal{F}^*$. Furthermore this projection is injective.
 \end{enumerate}
\end{Definition}
Let us first explain the meaning of the requirements that are present in this definition. The first two should be clear, we want an associative algebra that is on $\mathcal{F}$ given by the tensor $P_{abc}$, as is expected from pointwise defined algebras. The third condition above is introduced for two reasons. The first reason is physical, as we do not want low-energy functions to suddenly look differently once we are probing higher energies. Take for instance the unit function, we do not want to consider extensions of the algebra $\mathcal{F}$ where the unit of $\mathcal{F}$ does not correspond to the pointwise unit, as an associative extension might have a wildly different-looking dual space. Secondly, and this reason mainly refers to the injectivity requirement of the restriction, we do not want to consider extensions that are not connected to the current algebra through the product (for instance direct sums).

The associative extension is a first step towards the notion of the associative closure, and actually the associative closure we will define below is an associative extension. Consider for instance the algebra of smooth functions over a circle, from this definition it certainly is an associative extension of the five-dimensional algebra we considered around~\eqref{eq:duality:example_homo} as will be shown below. But the notion of associative extension does not have everything we want yet. For one, it is generally not unique for a tensor $P_{abc}$. Secondly, in the case of a finite-dimensional tensor that corresponds to a smooth manifold, there is an infinite-dimensional extension which we are looking for, but there are many finite-dimensional ones too. Thus we need a way to construct the infinite-dimensional one we are looking for (for instance the algebra of all smooth functions over a circle), if it exists.

Let us elaborate the example of the flat circle a bit more. The algebra we are after, $C^\infty\big(S^1\big)$, equipped with the $L^2\big(S^1\big)$ inner product is an algebraic extension. We can explicitly check this by going through the three requirements mentioned above. The first two points are relatively straightforward. Since the algebra is defined as a pointwise product, it is necessarily commutative and associative. Furthermore, the unit is a smooth function and thus included in this algebra making it a unital algebra. By the definition of the tensor in~\eqref{eq:def_P:innerp}, the algebra reduces to $P_{abc}$ when restricted to $\mathcal{F} \cong \mathbb{R}^5$. However, there are many other possible extensions of $P_{abc}$. As will be shown in Section~\ref{sec:associative_constr}, using the tensor rank decomposition it is possible to find a~seven-dimensional algebra with seven points in its algebraic dual space that is an associative extension of the five-dimensional $P_{abc}$ from the example. In fact, there are many of these associative extensions possible, and in a heuristic manner we can see the associative closure we want as a kind of ``union'' of those. It is however not useful to look at the union of the algebras themselves or their dual spaces, since the algebraic extensions are hardly comparable since they have different dimensions. What is comparable though are the extensions and their dual spaces when projected to $\mathcal{F}$ and $\mathcal{F}^*$, respectively. The restriction of an associative extension $\mathcal{A}$ to $\mathcal{F}$ just yields $\mathcal{A}|_{\mathcal{F}} = \mathcal{F}$ as it is an extension, thus we are left with considering the projection of the algebraic dual space~$|\mathcal{A}||_{\mathcal{F}^*}$.

From this, we are led to the following definition.
\begin{Definition}
 A \emph{potential homomorphism} of a tensor $P_{abc}$ acting on countable Hilbert space~$\mathcal{F}$ is a homomorphism of an associative extension $(\mathcal{A},\cdot)$ projected to $\mathcal{F}^*$. The \emph{space of potential homomorphisms} of $\mathcal{F}$, denoted $|\mathcal{F}|^{(P)}$, is the collection of all homomorphisms of all associative extensions projected to $\mathcal{F}^*$.
\end{Definition}
The key point of this definition is to allow us to not only consider the homomorphisms of $\mathcal{F}$ under $P_{abc}$, but consider all homomorphisms of some bigger algebra that could generate $P_{abc}$. We can think of this as the actual space of points that we are considering. This thought then finally leads us to the definition of the associative closure.
\begin{Definition}
 An associative closure of $\mathcal{F}$ is an associative extension $\mathcal{A}$ such that the restriction of its algebraic dual space to the linear dual space of $\mathcal{F}$ is exactly the space of potential homomorphisms of $\mathcal{F}$, i.e., $\left.|\mathcal{A}|\right|_{\mathcal{F}^*}=|\mathcal{F}|^{(P)}$.
\end{Definition}

\begin{Corollary}
The algebraic dual space of an algebraic closure is isomorphic to the space of potential homomorphisms, since the projection map is both injective and surjective.
\end{Corollary}
Note that while the associative closure is not necessarily unique, this is physically not really a problem. We assume here that what we can actually measure are the functions included in~$\mathcal{F}$, and the physical space is the space of potential homomorphisms. The associative closure is in that sense mainly a mathematical tool to be sure that there exists a topology on the space of potential homomorphisms. In the sections below we will explicitly construct an associative closure for finite-dimensional tensors, but it might be interesting to find out if the existence holds more generally. Furthermore, though this has not been investigated further, it might be the case that the different associative closures are related by some well-defined transformation, for instance a diffeomorphism.

As an example, let us show that the $C^\infty\big(S^1\big)$ algebra is an associative closure of the example of the flat circle.

\begin{Proposition}
 $C^\infty\big(S^1\big)$ is an associative closure of the five-dimensional tensor $P_{abc}$ discussed in the end of Section~$\ref{sec:topology}$.
\end{Proposition}
\begin{proof}
 As has already been argued before, $C^\infty\big(S^1\big)$ is an associative extension of $P_{abc}$. What we would like to show now is that every potential homomorphism of $P_{abc}$ is an element of $\big|C^\infty\big(S^1\big)\big|\big|_{\mathcal{F}^*}$.

 In this case, we can use the knowledge of the partial algebra $\mathcal{S} \cong \mathbb{R}^3$ mentioned above to our advantage. Firstly, the dual space of any covering system of partial algebra has to lie in the dual space of $\mathcal{S}$ since other systems are either equivalent, like $\mathcal{S}_\alpha$ mentioned above, or more restrictive. Thus, from the definition of any associative extension $\mathcal{A}$, the restriction of $|\mathcal{A}|$ to~$\mathcal{F}^*$ has to lie in $|\mathcal{S}|$, i.e., $|\mathcal{A}||_{\mathcal{F}^*} \subset |\mathcal{S}|$. In other words, every potential homomorphism must lie in~$|\mathcal{S}|$. This means that, if $\big|C^{\infty}\big(S^1\big)\big|\big|_{\mathcal{F}^*} = |\mathcal{S}|$, then necessarily every potential homomorphism lies in $\big|C^{\infty}\big(S^1\big)\big|\big|_{\mathcal{F}^*}$ and thus $C^\infty\big(S^1\big)$ is an associative closure.

 To do this, let us take an element $p\in|\mathcal{S}|$. From~\eqref{eq:duality:example_homo}, this is any $p\in\mathbb{R}^5$ such that $\forall a,b \leq 3$
 \begin{equation*}
 p_a p_b = \sum_{c=1}^5 P_{abc} p_c.
 \end{equation*}
 We can convince ourselves that for $p_1$, $p_2$, $p_3$ this already fixes the components as evaluation maps of $f_1 \sim 1$, $f_2 \sim \sin(x)$, $f_3 \sim \cos(x)$. This is because the evaluation maps of these functions are precisely all the homomorphisms $p \colon C^\infty\big(S^1\big) \rightarrow \mathbb{R}$, and since all the information of their products is contained within the five-dimensional tensor $P_{abc}$, the solutions to the above equation are nothing more but the restriction of these homomorphisms to $\mathbb{R}^3$, so $\big|C^{\infty}\big(S^1\big)\big|\big|_{\mathbb{R}^3}=|\mathcal{S}||_{\mathbb{R}^3}$. We now need to fix the last two elements and show that these correspond to restrictions of the full $C^\infty\big(S^1\big)$ homomorphisms.

 However, since the $C^\infty\big(S^1\big)$ algebra is defined pointwise, this is necessarily the case. For instance, let us take the case of $f_5$. From~\eqref{eq:duality:example_product},
 \begin{equation*}
 f_5 = \sqrt{2}f_1 - 2\sqrt{\pi} f_2 \cdot f_2 ,
 \end{equation*}
 and taking the same homomorphism $p \in \mathbb{R}^5$
 \begin{equation*}
 f_5(p) = p_5 = \sqrt{2} p_1 - 2\sqrt{\pi} p_2 p_2 = \sqrt{2} f_1(p) - 2\sqrt{\pi} f_2(p) f_2(p),
 \end{equation*}
 we get a pointwise definition of the function $f_5 \sim \cos(2x)$. A similar statement holds for $f_4 \sim \sin(2x)$. Since every homomorphism of $C^\infty\big(S^1\big)$ necessarily respects this, we see that actually
 \begin{equation*}
 \big|C^{\infty}\big(S^1\big)\big|\big|_{\mathcal{F}^*}=|\mathcal{S}|.
\tag*{\qed}
\end{equation*}
\renewcommand{\qed}{}
\end{proof}

A final note for this section is about the general structure of these definitions. While the algebra $C^\infty\big(S^1\big)$ is of course very nice in the sense that it exactly corresponds to a smooth manifold, these definitions can both reconstruct these nicer spaces or less ``nice'' spaces. The latter we will call fuzzy spaces, since as we will see later in Section~\ref{sec:cases}, their behaviour can vary from the usual continuous topological spaces.

\section{Constructing the associative closure using a measure}\label{sec:associative_constr}
In this section, we continue to understand the associative closure. We will first describe the appearance of a measure for a broad class of tensors, and subsequently describe a way to construct the associative closure for finite-dimensional tensors using the tensor rank decomposition. Along the way we will again refer to the example of the flat circle, where we reconstruct the full infinite-dimensional algebra and the measure on the dual space from the five-dimensional tensor constructed in Section~\ref{sec:topology}. Note that in this section, we still assume the tensor $P_{abc}$ to be unital as explained in Section~\ref{sec:associative}, but not associative.

Generally, the associative closure of $P_{abc}$ acting on $\mathcal{F}$ may be constructed in two steps: 1)~generate the space of potential homomorphisms $|\mathcal{F}|^{(P)}$, 2)~construct the associative closure by treating the potential homomorphisms as evaluation maps of functions and taking pointwise products between them to generate new functions. The result of this will then be a new algebra that is pointwise defined, and thus associative by construction, with an algebraic dual space isomorphic to $|\mathcal{F}|^{(P)}$. In order to simplify the notation a bit, in this section we will assume~$P_{abc}$ to be a finite-dimensional tensor. We will first introduce the notion of measure generated tensors, after which we will show that these tensors have a natural associative closure. After this, we will show how to actually construct the space of potential homomorphisms and find such a~measure, where we will also state a curious conjecture: every unital tensor is measure-generated. Along the way, we will further develop the example of the flat circle mentioned in Section~\ref{sec:topology}.

\subsection{Measure generated tensors}\label{sec:associative_constr:measure}
In this section, we introduce the notion of measure generated tensors. This is a broad class of tensors, that roughly speaking can be interpreted as constructed using a set of points with~measure on them. The precise definition is as follows.\footnote{For some of the notions we use from measure theory, we would like to refer to Appendix~\ref{sec:app:measure}.}
\begin{Definition}
A tensor $P_{abc}$ acting on $\mathcal{F}$ is said to be \emph{measure-generated} if there is an associative closure $\mathcal{A}$ equipped with a measure $\mu$ on $|\mathcal{A}|$ such that
\begin{equation}\label{eq:associative_constr_measure_innerp}
 \forall f,g \in \mathcal{A}\colon \braket{f | g} = \int_{|\mathcal{A}|} {\rm d}\mu(p) f(p) g(p).
\end{equation}
\end{Definition}
Note that $|\mathcal{A}|$ comes with a natural topology as explained in Figure~\ref{fig:duality:topology}, thus we can construct the Borel $\sigma$-algebra on this space, on which the measure is defined (see Appendix~\ref{sec:app:measure}).
\begin{Proposition}
 For any $f,g\in\mathcal{F}$,
 \begin{equation*}
 \braket{f| g} = \int_{|\mathcal{A}|} {\rm d}\mu(p) f(p) g(p) = \int_{|\mathcal{F}|^{(P)}} {\rm d}\mu(p) f(p) g(p),
 \end{equation*}
\end{Proposition}
\begin{proof}
 $|\mathcal{F}|^{(P)}$ is the projection of homomorphisms $p\in |\mathcal{A}|$ to $\mathcal{F}^*$ by the definition of the associative closure, so for $f \in \mathcal{F}, p(f) = p^*(f)$, where $p^*$ denotes the projection of $p$ to $\mathcal{F}^*$. Since we are only considering functions in $\mathcal{F}$, and the spaces are isomorphic, we only have to consider~$|\mathcal{F}|^{(P)}$.
\end{proof}

\begin{Corollary}
 In particular, for any two basis elements $f_a, f_b \in \mathcal{B}_\mathcal{F}$,
 \begin{equation*}
 \delta_{ab} = \int_{|\mathcal{F}|^{(P)}} {\rm d}\mu(p) f_a(p) f_b(p).
 \end{equation*}
\end{Corollary}
\begin{Proposition}
A measure-generated unital tensor is given by
\begin{equation}\label{eq:associative_constr_measure_P}
 P_{abc} = \int_{|\mathcal{F}|^{(P)}} {\rm d}\mu(p) f_a(p) f_b(p) f_c(p).
\end{equation}
\end{Proposition}
\begin{proof}
 Take $\mathcal{A}$ to be an associative closure of $P_{abc}$. For an element of its algebraic dual space $p\in|\mathcal{A}|$ we have
 \begin{equation*}
 p(f_a) p(f_b) = p(f_a \cdot f_b).
 \end{equation*}
 Furthermore, for any product of the basis function $f_a, f_b \in \mathcal{B}_\mathcal{F}$,
 \begin{equation*}
 f_a \cdot f_b = P_{abc} f_c + g,
 \end{equation*}
 where $ P_{abc} f_c \in \mathcal{F}$ and $g\in \mathcal{A} \setminus \mathcal{F}$. This comes from the definition of an associative extension, such that $P_{abc}=\braket{f_c| f_a \cdot f_b}$, so $\braket{f| g} = 0$ for all $f \in \mathcal{F}$. In particular, for any $p\in|\mathcal{A}|$,
 \begin{equation*}
 p(f_a) p(f_b) = p(f_a \cdot f_b) = p(P_{abc} f_c) + p(g).
 \end{equation*}
 Since $p(f)=p^*(f)$ for $f\in\mathcal{F}$, where $p^*$ denotes the projection of $p$ to $\mathcal{F}^*$, we find
 \begin{align*}
 \int_{|\mathcal{F}|^{(P)}}{\rm d}\mu(p) p(f_a) p(f_b) p(f_c) &= \int_{|\mathcal{A}|}{\rm d}\mu(p) p(f_a) p(f_b) p(f_c),\\
 &= \int_{|\mathcal{A}|}{\rm d}\mu(p) p(P_{abd}f_d) p(f_c) + \int_{|\mathcal{A}|}{\rm d}\mu(p) p(g) p(f_c).
 \end{align*}
 On the right-hand side, we have the inner product as defined in~\eqref{eq:associative_constr_measure_innerp}, so we get
 \begin{equation*}
 \int_{|\mathcal{F}|^{(P)}}{\rm d}\mu(p) p(f_a) p(f_b) p(f_c) = P_{abd} \braket{f_d | f_c} + \braket{g| f_c} = P_{abc} ,
 \end{equation*}
 where the last inner product is zero since $g \not\in \mathcal{F}$.
\end{proof}

In the example of the flat circle, it is clear that the above definition is satisfied. First, the inner product on the smooth functions of the flat circle is given by~\eqref{eq:topology:innerp}, this means that we can interpret the measure above as the canonical Riemannian measure. The equation~\eqref{eq:associative_constr_measure_P} follows from~\eqref{eq:def_P:innerp}.

Let us develop some more intuition for these notions before continuing. Let us first assume that the basis functions are given by simple functions\footnote{See Appendix~\ref{sec:app:measure}.}
\begin{equation}\label{eq:associative_constr:simple_f}
 f_a = \sum_{i=1}^R p^i_a \mathbf{1}_{A_i},
\end{equation}
where $R$ is the amount of disjoint regions $A_i\subset |\mathcal{F}|^{(P)}$, $p^i_a$ denotes the value of the function $f_a$ in region $A_i$, and $\mathbf{1}_{A_i}$ is the region's indicator function. Given a measure on $|\mathcal{F}|^{(P)}$ we then find
\begin{equation*}
 \int_{|\mathcal{F}|^{(P)}} {\rm d}\mu(p) f_a = \sum_{i=1}^R p^i_a \mu(A_i) \equiv \sum_{i=1}^R p^i_a \beta_i.
\end{equation*}
Here $\beta_i \equiv \mu(A_i) > 0$. Note that we assume that all of the basis functions have the same decomposition in terms of the indicator functions. This means that if we evaluate a product we get a similar decomposition
\begin{equation*}
 f_a\cdot f_b = \sum_{i=1}^R p^i_a p^i_b \mathbf{1}_{A_i},
\end{equation*}
and, by using the definition above in~\eqref{eq:associative_constr_measure_innerp},
\begin{equation}\label{eq:associative_closure:simple_innerp}
 \braket{f_a | f_b} = \int_{|\mathcal{F}|^{(P)}} {\rm d}\mu(p) f_a(p) f_b(p) = \sum_{i=1}^R \mu(A_i) p^i_a p^i_b \equiv \sum_{i=1}^R \beta_i p^i_a p^i_b.
\end{equation}
Moreover, by using~\eqref{eq:associative_constr_measure_P} we find
\begin{equation}\label{eq:associative_closure:simple_tensor}
 P_{abc} = \sum_{i=1}^R \beta_i p^i_a p^i_b p^i_c.
\end{equation}
This expression may be recognised as a tensor rank decomposition of the tensor $P_{abc}$, and it is a first sign of why the tensor rank decomposition is so useful and has been so successful when applied to the canonical tensor model in the past. Usually the basis functions of an algebra of functions over a space are not given by simple functions, but a finite-dimensional part of the algebra may still be represented by them. In Section~\ref{sec:associative_constr:pot_hom}, we will see that the tensor rank decomposition actually generates a set of $p^i_a$ and $\beta_i$ which represents the algebra in a~pointwise manner, and actually corresponds to an associative extension. We will argue that taking all possible tensor rank decompositions (under certain restrictions) will correspond to an associative closure.

\begin{figure}
 \centering
 \begin{minipage}{0.3\textwidth}
 \includegraphics[width=0.95\textwidth]{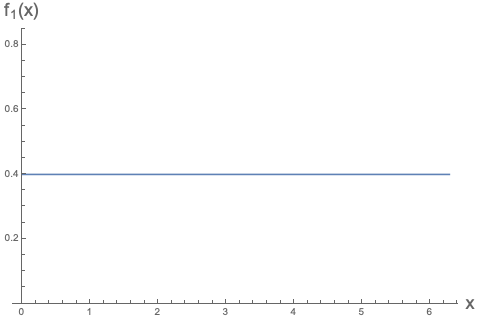}
 \end{minipage}
 \begin{minipage}{0.3\textwidth}
 \includegraphics[width=0.95\textwidth]{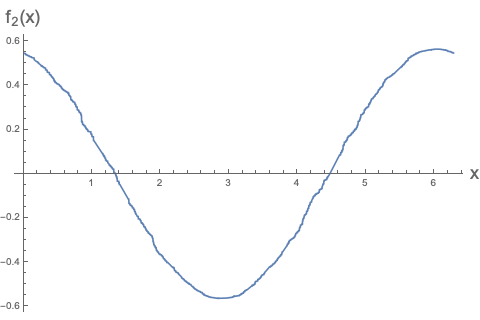}
 \end{minipage}
 \begin{minipage}{0.3\textwidth}
 \includegraphics[width=0.95\textwidth]{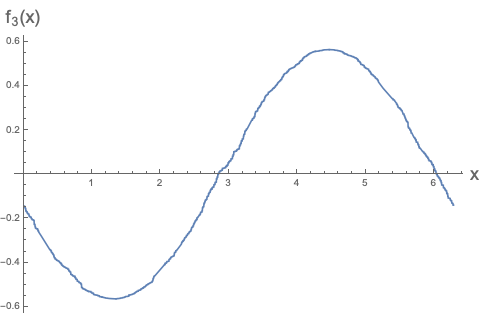}
 \end{minipage}

 \caption{A plot of the first three basis functions of the circle, as given by~\eqref{eq:associative_closure:measure:f_representation}. As mentioned in the text, $1000$ points are used. Note that the deformations of Figure~\ref{fig:duality:circle_topology} disappear due to the inclusion of the information of the measure.}
 \label{fig:ass_constr:circle_measure}
\end{figure}

Going back to the example of the flat circle, we took $R=1000$ points of the $10000$ points calculated in Figure~\ref{fig:duality:circle_topology} at random. We then used \textsc{Mathematica} to find a solution for~\eqref{eq:associative_closure:simple_tensor}. In the case of the circle, the $\beta_i$ can be interpreted as the length of each line-segment since line-segments generate the measurable sets of the circle and $\beta_i=\mu(A_i)$ is their length. This implies that we can approximate the functions $f_a$ as simple functions
\begin{equation}\label{eq:associative_closure:measure:f_representation}
 f_a(x) = \begin{cases}
 p_a^1 , & 0\leq x < \beta_1,\\
 p_a^2 , & \beta_1\leq x < \beta_1 + \beta_2,\\
 \dots & \\
 p_a^R , & \displaystyle \sum_{i=1}^{R-1} \beta_i \leq x < \sum_{i=1}^{R} \beta_i.
 \end{cases}
\end{equation}
The result of this procedure may be found in Figure~\ref{fig:ass_constr:circle_measure}. Note that the circumference of the circle (the range of the plots) exactly matches $2\pi$, and the functions look exactly like the functions we started out with in the construction of the tensor in~\eqref{eq:topology:example:basis}. It should be noted that this is merely an approximation of the functions, but in the limit they will exactly represent the original functions $\{1, \sin(x), \cos(x)\}$.

While the solution in Figure~\ref{fig:ass_constr:circle_measure} looks exactly like the functions we are after, it is not a~definition of the full measure yet by the $\beta_i$. To reconstruct the full measure, one needs to take all possible finite subsets of $|\mathcal{F}|^{(P)}$, and take all possible solutions to~\eqref{eq:associative_closure:simple_innerp} (or equivalently~\eqref{eq:associative_closure:simple_tensor}). The collection of all these different solutions will then correspond to the measure over different regions. In terms of measure theory, this would allow us to then integrate any function using Lebesgue integration since we know the measure-value of the indicator functions of any region.

Note that in practice, it usually suffices to take only one set of points and find a solution like in Figure~\ref{fig:ass_constr:circle_measure}, similarly to the fact that for many purposes taking a finite Riemann sum gives a good approximation of many integrals. In Section~\ref{sec:associative_constr:constr}, we will show that the $1000$ points evaluated here can very accurately generate new basis elements, at least up to $N=11$.

The notion of a measure-generated tensor is important here, and one might be worried that this limits the scope of this section considerably. However, in Section~\ref{sec:associative_constr:pot_hom}, we will argue that many~-- if not all~-- unital tensors actually have this property.

\subsection{Construction of an associative closure}\label{sec:associative_constr:constr}
In this section, we will show how to construct an associative closure from a measure-generated tensor. We will assume that one already constructed the space of potential homomorphisms. In Section~\ref{sec:associative_constr:pot_hom}, we will show how one could do this.

Let us take two basis functions $f_a, f_b \in \mathcal{B}_\mathcal{F}$, and consider a potential homomorphism ${p\!\in\!|\mathcal{F}|^{(P)}}$ and treat it as a point, i.e., evaluation map. The pointwise product of these functions can then be decomposed as
\begin{equation}\label{eq:associative_closure:constr:start}
 f_a(p) f_b(p) = P_{abc}f_c(p) + g(p) = \braket{f_c | f_a \cdot f_b} f_c(p) + g(p).
\end{equation}
Here $g(p)$ corresponds to the difference between the pointwise product to the product induced by $P_{abc}$. If $g(p)=0$ it means that the tensor already describes a pointwise algebra for $f_a$, $f_b$,\footnote{As an aside, this also implies that there is an associative commutative partial algebra spanned by $f_a$, $f_b$.} and we try another combination of basis functions such that $g(p)\neq 0$ for them. If there is no such combination, the algebra described by $P_{abc}$ is already a pointwise (associative) algebra and we are done. From here on we assume that $g(p)\neq 0$.

The general idea is now that we see $g(p)$ as a new function that is not in our algebra yet. By construction, this $g$ is a map
\begin{equation*}
 g \colon \ |\mathcal{F}|^{(P)} \rightarrow \mathbb{R}.
\end{equation*}
In a sense we ``close'' the algebra by adding $g$ as a new basis element. Note that by construction it is true that $\forall f_a \in \mathcal{B}_{\mathcal{F}}$
\begin{equation*}
 \int_{|\mathcal{F}|^{(P)}} {\rm d}\mu(p) f_a(p) g(p) = 0.
\end{equation*}
This means that actually we can extend the inner product to $g$ as well. Say the dimension of $\mathcal{F}$ is $N$, then we can now add a new basis element
\begin{equation*}
 f_{N+1}(p) = \frac{g(p)}{\|g\|},
\end{equation*}
where $\|g\| = \sqrt{\braket{g| g}}$ is the norm induced by the inner product. $f_{N+1}$ together with $\mathcal{B}_\mathcal{F}$ forms an orthonormal basis with the same inner product
\begin{equation}\label{eq:associative_closure:constr:new_innerp}
 \delta_{ab} = \braket{f_a | f_b} = \int_{|\mathcal{F}|^{(P)}} {\rm d}\mu(p) f_a(p) f_b(p),
\end{equation}
where $a$ now runs from $1$ to $N+1$. Let us now denote this vector space by $F^{(N+1)}$, and introduce a new tensor $P_{abc}^{(N+1)} = \braket{f_c| f_a \cdot f_b}$ for all $a,b,c \leq N+1$ with the inner product of~\eqref{eq:associative_closure:constr:new_innerp}.

Next we pick new $f_a, f_b \in \mathcal{F}^{(N+1)}$ and restart this process starting with~\eqref{eq:associative_closure:constr:start}, where we treat $\mathcal{F}^{(N+1)}$ as the vector space with tensor $P_{abc}^{(N+1)}$. When we keep iterating, we will either eventually reach a point where for every $f_a$, $f_b$ the product is exactly described by $P_{abc}$, i.e., $f_a(p) f_b(p) = P_{abc} f_c(p)$, or we continue finding new functions. In the former, at that point we have a proper definition of the associative closure, which will be finite-dimensional. In the latter case, we define the associative closure as the inductive limit of this process, producing a~countably infinite-dimensional algebra.

Let us now explicitly check that the algebra constructed like this is indeed an associative closure. For this, we first need to confirm that it is an associative extension. First note that for every element in $p^*\in|\mathcal{F}|^{(P)}$ we constructed an element in the algebraic dual space of $\mathcal{A}$, $p\in|\mathcal{A}|$, $p(f_a) = f_a(p^*)$, since we basically constructed a new algebra where the original potential homomorphisms become proper homomorphisms. Referring to Definition~\ref{def:associative:ass_extension}, it is clear that the first two points are satisfied by construction. Furthermore, since the every element of the dual space $p\in |\mathcal{A}|$ projected to $\mathcal{F}^*$ must be an element of some associative extension by the definition of potential homomorphisms, it necessarily lies in $|\mathcal{S}|$ for a maximal system of partial algebras $\mathcal{S}_i\subset\mathcal{F}$. Since every element $p\in|\mathcal{A}|$ is constructed from a $p^*\in|\mathcal{F}|^{(P)}$, this is necessarily injective. Similarly, it follows that it is an associative closure, since for every $p\in|\mathcal{A}|$ it must hold that $p^* \equiv p|_{\mathcal{F}^*}\in|\mathcal{F}|^{(P)}$ by definition of the space of potential homomorphisms, and we constructed a $p\in|\mathcal{A}|$ from every $p^* \in |\mathcal{F}|^{(P)}$, we have an isomorphism and thus $|\mathcal{A}||_{\mathcal{F}^*} = |\mathcal{F}|^{(P)}$.

\begin{figure}
 \centering
 \begin{minipage}{0.45\textwidth}
 \includegraphics[width=0.95\textwidth]{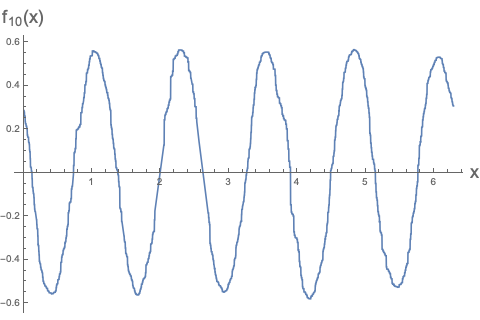}
 \end{minipage}
 \begin{minipage}{0.45\textwidth}
 \includegraphics[width=0.95\textwidth]{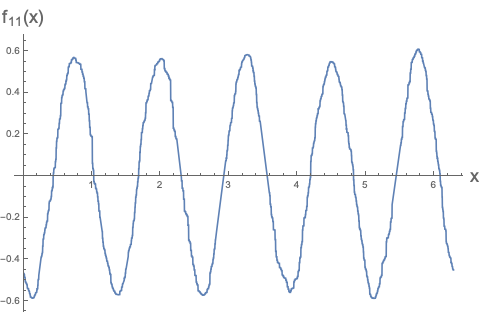}
 \end{minipage}

 \caption{A plot of the 10th and 11th basis functions of the flat circle generated as described by the procedure in Section~\ref{sec:associative_constr:constr} and using the simple function inner product of~\eqref{eq:associative_closure:simple_innerp} with the same $1000$ points as before in Figure~\ref{fig:ass_constr:circle_measure}.}
 \label{fig:ass_constr:circle_closure}
\end{figure}
Let us now go back to the flat circle. As we already know that the associative closure $\mathcal{A}\cong C^{\infty}\big(S^1\big)$ is infinite-dimensional, this procedure would never terminate. We used this construction together with the simple function approximation introduced in Section~\ref{sec:associative_constr:measure} to calculate the inner product in order to extend the algebra $\mathcal{F}\cong \mathbb{R}^5$ to an extension $\mathcal{A}\cong \mathbb{R}^{11}$. The result for $f_{10}(x) \sim \sin(5x)$ and $f_{11}(x) \sim \cos(5x)$ (up to some phase due to the random choice of the first point) is given in Figure~\ref{fig:ass_constr:circle_closure}. Note that technically the construction with a finite amount of points is an associative extension and not a closure, but using more and more points would create a~more and more accurate representation of the functions up to very high dimensions and in the limit the associative closure.

\subsection{Generating potential homomorphisms}\label{sec:associative_constr:pot_hom}
In this section, we show how to generate the potential homomorphisms more generally. In the previous sections, we always assumed that we knew the potential homomorphisms of an algebra already, and in the case of the flat circle we ``got lucky'' since we could simply take the partial algebra homomorphisms as explained in~\eqref{eq:duality:example_homo} and properly defined in Definition~\ref{def:associative:dual_space}. In this section, we will use an important tool that has been extensively used in the context of the canonical tensor model~\cite{Kawano:2018pip}, the tensor rank decomposition. We will end this section by again examining the example of the flat circle.

Let us consider a tensor rank decomposition for some (not necessarily the lowest) $R$ of the unital tensor $P_{abc}$\footnote{Note that we assume a tensor rank decomposition to be any decomposition of $P_{abc}$ such that~\eqref{eq:associative_constr:TRD_phi} is satisfied as in~\cite{Obster:2021xtb}. Oftentimes it is defined instead to be the decomposition for the lowest $R$ possible, which we call a~\emph{minimal} tensor rank decomposition.}
\begin{equation}\label{eq:associative_constr:TRD_phi}
 P_{abc} = \sum_{i=1}^R \phi^i_a \phi^i_b \phi^i_c.
\end{equation}
Since the tensor algebra is unital, we know for the unit $1=\alpha^a f_a$,
\begin{equation}\label{eq:associative_constr:TRD_phi_innerp}
 \delta_{bc} = \alpha^a P_{abc} = \sum_{i=1}^R (\alpha^a \phi^i_a) \phi^i_b \phi^i_c \equiv \sum_{i=1}^R \gamma_i \phi^i_b \phi^i_c.
\end{equation}
We call the tensor rank decomposition \emph{positive} if $\gamma_i > 0$ for all $i$. If we now define $\beta_i = (\gamma_i)^{1/3} > 0$ and $p^i_a \equiv (\beta_i)^{-1/3} \phi^i_a$, we see that~\eqref{eq:associative_constr:TRD_phi} and~\eqref{eq:associative_constr:TRD_phi_innerp} exactly reduce to
\begin{gather}\label{eq:associative_constr:TRD_betappp}
 P_{abc} = \sum_{i=1}^R \beta_i p_a^i p_b^i p_c^i,\qquad
 \delta_{ab} = \sum_{i=1}^R \beta_i p_a^i p_b^i,
\end{gather}
in a similar fashion to~\eqref{eq:associative_closure:simple_tensor} and~\eqref{eq:associative_closure:simple_innerp} with $\beta_i>0$.

The similarity between~\eqref{eq:associative_constr:TRD_betappp} and \eqref{eq:associative_closure:simple_tensor} suggests that the tensor rank decomposition has a relationship with the simple function approach using potential homomorphisms of Section~\ref{sec:associative_constr:measure}. The idea in the following is to treat the $p_a^i$ above as potential homomorphisms. If the $p^i\in \mathcal{F}^*$ indeed correspond to potential homomorphisms $p^i \in |\mathcal{F}|^{(P)} \subset \mathcal{F}^*$ we could keep finding tensor rank decompositions in order to generate more and more points. Furthermore, if we consider the interpretation introduced in Section~\ref{sec:associative_constr:measure}, we already get the information of the measure for free through the $\beta_i$. Then, using the procedure described in Section~\ref{sec:associative_constr:constr} we can reconstruct the whole associative closure.

\begin{Definition}
 A positive tensor rank decomposition $P_{abc} = \sum_{i=1}^R \beta_i p_a^i p_b^i p_c^i$ (note, not necessarily minimal) is called a \emph{pointwise decomposition} if every $p_a^i\in |\mathcal{F}|^{(P)}$.
\end{Definition}
By this definition, the decomposition in~\eqref{eq:associative_closure:simple_tensor} is a pointwise decomposition. Thus we immediately can conclude that
\begin{Corollary}
 Every measure-generated tensor has a pointwise decomposition.
\end{Corollary}

Naturally, this bring up several questions. Firstly, whether or not every tensor have a positive tensor rank decomposition. There is no conclusive proof of this fact yet, but some numerical calculations for $N=3,4,5$ imply that it seems to be true. The calculation was done using \textsc{Mathematica} and generating $1000$ random unital symmetric tensors $P_{abc}$ and trying to obtain a positive tensor rank decomposition. The tensors were ensured to be unital by fixing $P_{1ab} = P_{a1b} = P_{ab1} = \delta_{ab}$. In all cases a good decomposition with an error $\big(\sum_{i=1}^R \beta_i p_a^i p_b^i p_c^i - P_{abc}\big)^2 < 10^{-20}$ could be found. This leads to the formulation of the conjecture:
\begin{Conjecture}
 Every unital symmetric real tensor has a positive tensor rank decomposition.
\end{Conjecture}
Though the numerical evidence suggests this to be true, it still remains to be proven.

Secondly, if we have such a positive tensor rank decomposition, will it be a pointwise decomposition? The answer to this question in general is no. Note that in the above requirement that $\gamma^i > 0$ is similar to requiring that the unit of $P_{abc}$ becomes a pointwise unit. It does not necessarily mean that all other functions also suddenly have proper pointwise representations. This seems to make it necessary to check every time if indeed all of the elements of the tensor rank decomposition $p^i$ correspond to potential homomorphisms. However, the example of the flat circle that is discussed below suggests that it seems to be the case for the \emph{minimal} positive tensor rank decomposition. This leads to a second conjecture:
\begin{Conjecture}
 For a unital tensor $P_{abc}$ that admits a covering system of partial algebras, a~minimal positive tensor rank decomposition is a pointwise decomposition.
\end{Conjecture}
It is at the moment not known how to prove this statement, and whether it is applicable even more generally. However, if the tensor does not admit a covering system of partial algebras, it seems necessary to actually verify if the $p^i$ are potential homomorphisms. In practice, this is done by checking if they are in the dual space of a maximal system of partial algebras. This is also a spot where the algebraic approach and the pure tensor rank decomposition approach might not always agree, which might also have to do with the tensors in the quantum- and geometric phases of the canonical tensor model~\cite{Kawano:2021vqc}. In the future, it might prove beneficial to either restrict the tensor rank decomposition approach further, or broaden the definition of the associative extension.

Once one finds the collection $p^i$ as described above, it is then possible to construct an associative extension, and with all possible $p^i$ a closure, of the tensor $P_{abc}$ by using the method described in Section~\ref{sec:associative_constr:constr}.

Using what is written above, we can then interpret the positive tensor rank decomposition as a simple functions representation of $f\in\mathcal{F}$, as introduced in~\eqref{eq:associative_constr:simple_f}. As was already pointed out in~\cite{Kawano:2021vqc}, if the tensor $P_{abc}$ has a symmetry such that for a Lie-group transformation $G_{ab}$
\begin{equation*}
 G_{aa'}G_{bb'}G_{cc'} P_{a'b'c'} = P_{abc},
\end{equation*}
there is a continuous degeneracy of the tensor rank decomposition, since for every $p^i_a$ belonging to a tensor rank decomposition, $G_{aa'}p^i_{a'}$ belongs to an element of another tensor rank decomposition%
\begin{equation*}
 G_{aa'}G_{bb'}G_{cc'} P_{a'b'c'} = \sum_{i=1}^R \beta_i \big(G_{aa'} p_{a'}^i\big) \big(G_{bb'} p_{b'}^i\big) \big(G_{cc'} p_{c'}^i\big) = P_{abc}.
\end{equation*}
This is the reason that we can only say that a tensor rank decomposition corresponds to an associative extension, since not all potential homomorphisms are generally included. Once we consider all pointwise decompositions, we will get the whole space of potential homomorphisms~$|\mathcal{F}|^{(P)}$, which we can then use to construct the full associative closure as in Section~\ref{sec:associative_constr:constr}. In~\cite{Obster:2017pdq, Obster:2017dhx}, it was already argued that the canonical tensor model seems to prefer symmetric states, with the current discussion this implies that we can expect the emergence of (almost) continuous spacetimes.

\begin{figure}[t]
 \centering
 \begin{minipage}{0.3\textwidth}
 \includegraphics[width=0.95\textwidth]{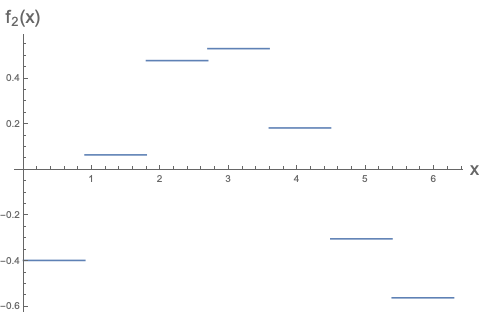}
 \end{minipage}
 \begin{minipage}{0.3\textwidth}
 \includegraphics[width=0.95\textwidth]{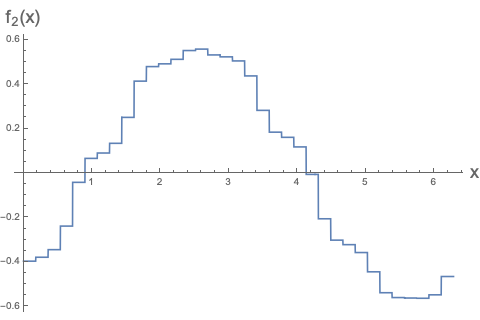}
 \end{minipage}
 \begin{minipage}{0.3\textwidth}
 \includegraphics[width=0.95\textwidth]{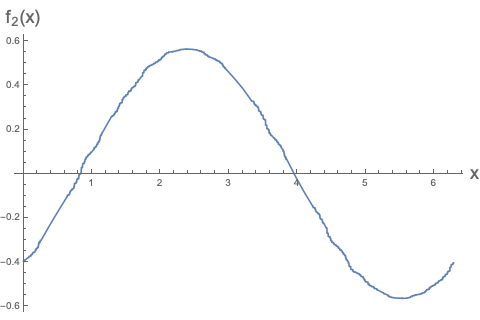}
 \end{minipage}

 \caption{A plot of the potential homomorphisms generated by the tensor rank decomposition of the second basis function of the flat circle, using the representation in~\eqref{eq:associative_closure:measure:f_representation}. As mentioned in the text, we used from left to right 1, 5 and 150 tensor rank decompositions, leading to 7, 35 and 1050 points, respectively. Note the agreement between the rightmost figure and Figure~\ref{fig:ass_constr:circle_measure} (except for a phase difference due to the random choice of the first point).} \label{fig:ass_constr:circle_TRD}
\end{figure}

Before finishing this section, we will examine the flat circle once more. The tensor described in~\eqref{eq:duality:example_product} has rank $R=7$ which can be verified by finding solutions to $0 = \big(\sum_{i=1}^R \phi_a^i \phi_b^i \phi_c^i - P_{abc}\big)^2$. In the following, we denote by $M$ the amount of tensor rank decompositions generated. We have found $M=150$ minimal tensor rank decompositions using \textsc{Mathematica} and treated the points as defined in~\eqref{eq:associative_constr:TRD_betappp}. As expected from the discussion, these points $\{p^i\}$ appear to be potential homomorphisms since they satisfy
\begin{equation}\label{eq:associative:trd:check}
 s_1^a s_2^b p_a^i p_b^i = s_1^a s_2^b P_{abc} p_c^i
\end{equation}
for every $s_1, s_2 \in \mathcal{S}_3$, the covering partial algebra as defined in Section~\ref{sec:associative}. Moreover; we seem to find all possible homomorphisms. After generating $M$ minimal tensor rank decompositions, the $I$-th one denoted by $\sum_{i=1}^R \beta_i^{(I)} {p_a^{(I)}}^i {p_b^{(I)}}^i {p_c^{(I)}}^i $, we sum them together as
\begin{align}
 P_{abc} &= \frac{1}{M} \Bigg( \sum_{i=1}^R \beta_{i}^{(1)} {p_a^{(1)}}^{i} {p_b^{(1)}}^{i}{p_c^{(1)}}^{i} + \dots + \sum_{i_M=1}^R \beta_{i}^{(M)} {p_a^{(M)}}^{i}{p_b^{(M)}}^{i}{p_c^{(M)}}^{i} \Bigg), \nonumber\\
 &= \frac{1}{M} \sum_{I=1}^M \sum_{i=1}^R \beta^{(I)}_{i} {p_a^{(I)}}^i {p_b^{(I)}}^i {p_c^{(I)}}^i.\label{eq:associative_closure:trd:sum}
\end{align}
Using the same simple function representation as in~\eqref{eq:associative_closure:measure:f_representation}, we plotted the values of $f_2(x)$ using the same ordering as discussed at~\eqref{eq:duality:example:topology_norm}. We did this for three cases: using only $M=1$ tensor rank decomposition, using $M=5$ tensor rank decompositions and using $M=150$ tensor rank decompositions.

We can now view the sum in~\eqref{eq:associative_closure:trd:sum} as one big tensor rank decomposition, consisting of $M*R$ points. For the measure-values $\beta_i$ we need to make sure that the sum $\sum_{i=1}^{M*R} \beta_i = \sum_{i=1}^{R} \beta^{(I)}_i \ \forall_{I \leq M}$, as otherwise we would get a tensor rank decomposition for $M*P_{abc}$. In the above, we used the rule, for $I\leq M,$ and $j\leq R$
\begin{equation}\label{eq:associative_constr:beta_example}
 \beta_{(I-1)*R + j} = \frac{1}{M} \beta^{(I)}_{j}.
\end{equation}
There is a certain ambiguity in doing so. Making a different choice for any
\begin{equation*}
 \beta_{(I-1)*R+j} = A_{I}\beta^{(I)}_{j},
\end{equation*}
such that $A_I > 0$, $\sum_{I=1}^M A_I = 1$, and technically this would correspond to a different associative closure. It is expected that such a ``transformation'' corresponds to performing a diffeomorphism on the manifold, since it can be interpreted as a deformation. Here we chose simply~\eqref{eq:associative_constr:beta_example} and the result of this exercise may be found in Figure~\ref{fig:ass_constr:circle_TRD}.

\begin{figure}
 \centering
 \begin{minipage}{0.3\textwidth}
 \includegraphics[width=0.95\textwidth]{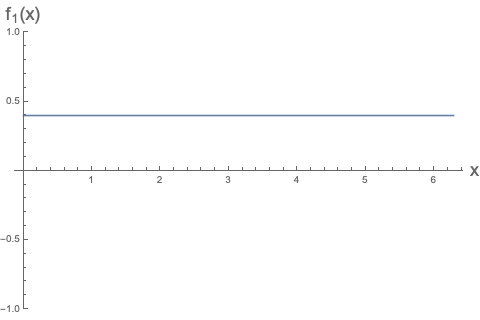}
 \end{minipage}
 \begin{minipage}{0.3\textwidth}
 \includegraphics[width=0.95\textwidth]{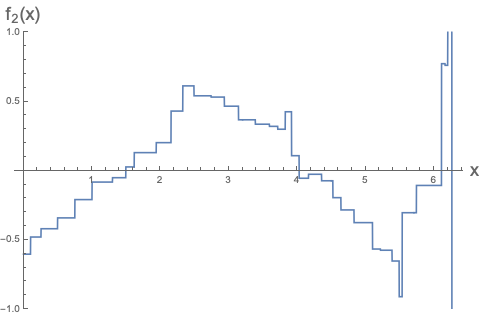}
 \end{minipage}
 \begin{minipage}{0.3\textwidth}
 \includegraphics[width=0.95\textwidth]{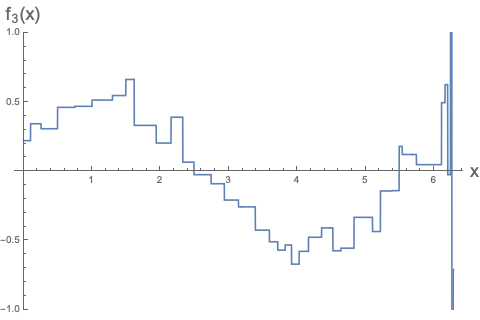}
 \end{minipage}

 \caption{A plot of the elements generated by a non-minimal tensor rank decomposition, showing how they deviate from the results in Figure~\ref{fig:ass_constr:circle_TRD}. In this image, $10$ tensor rank decompositions of rank $8$ are combined into the plot, using the same procedure as before. Some points still look similar to Figure~\ref{fig:ass_constr:circle_TRD}, but there are clear discrepancies from what we are looking for.}
 \label{fig:ass_constr:r=8}
\end{figure}

We end this section with a remark about the requirement of the tensor rank decomposition to be \emph{minimal}. It might not seem clear from the discussion above why this should be the case, but the empirical evidence points towards this. Let us consider the example of the flat circle cited above again. If we attempt to find $R=8$ tensor rank decompositions, instead of the minimal $R=7$ above, we see that the resulting tensor rank decomposition does not consist of potential homomorphisms. This is shown visually in Figure~\ref{fig:ass_constr:r=8}, but can also be verified by verifying~\eqref{eq:associative:trd:check}. For the case of $R=7$, all of the elements $p^i_a$ in the tensor rank decomposition exactly satisfied~\eqref{eq:associative:trd:check}, but for $R=8$ there always seem to be two elements $p_a^i$ that deviate a~lot \big(as in $\big|s_1^a s_2^b p_a^i p_b^i - s_1^a s_2^b P_{abc} p_c^i\big| \sim O(1)$\big) and the other six deviate a little (where the deviation was ${\sim} O(0.1)$), explaining why there seem to be some random points in Figure~\ref{fig:ass_constr:r=8} along with points that look very similar to Figure~\ref{fig:ass_constr:circle_TRD}. It should be noted that the $f_1$ function, proportional to the unit function, does always yield the correct result.

\section{Defining a unital algebra and introducing geometry}\label{sec:unit}
Up until now, we have always assumed that the tensors we used already have a unit, in the sense that there is a $\mathbf{1} = \alpha_a f_a \in \mathcal{F}$ such that
\begin{equation}\label{eq:unit:def_unit}
 \alpha_a P_{abc} = \delta_{bc}.
\end{equation}
For a general symmetric tensor $P_{abc}$, there is no such element $\mathbf{1}$. In order to fully interpret these tensors as representing algebras of functions over a compact Riemannian manifold, the algebra needs to contain a unit. In this section, we introduce a way to generate a new tensor $\tilde{P}_{abc}$ which does have a unit, which is possible to do with a large class of tensors $P_{abc}$ as will be explained below. The method used will have some important physical implications which are discussed towards the end of the section, since this gives us the opportunity to describe the full geometric information of a Riemannian manifold in the tensor $P_{abc}$ by interpreting the deviation of the unit as the eigenvalues of the Laplace--Beltrami operator.

First let us note that the unit in~\eqref{eq:unit:def_unit} was derived from the requirement that $\forall f = \beta_a f_a \in \mathcal{F}$%
\begin{equation*}
 \mathbf{1} \cdot f = \alpha_a \beta_b P_{abc} f_c = \beta_c f_c = f.
\end{equation*}
This notion might be generalised to the requirement that the unit is its own unit
\begin{equation*}
 \mathbf{1} \cdot \mathbf{1} = \alpha_a \alpha_b P_{abc} f_c = \alpha_c f_c = \mathbf{1},
\end{equation*}
which implies
\begin{equation}\label{eq:unit:gen_unit}
 \alpha_a \alpha_b P_{abc} = \alpha_c.
\end{equation}
This might be recognised as the eigen-problem of a tensor $P_{abc}$. For a real tensor, there is always at least one real solution to this~\cite{QI20051302}. If there are several real solutions, one has to make a choice. We will get back to this later in the section.

After finding an $\alpha_a$ such that~\eqref{eq:unit:gen_unit} holds, we consider the matrix
\begin{equation*}
 M_{bc} = \alpha_a P_{abc}.
\end{equation*}
If $\alpha_a$ would correspond to a true unit, this matrix would simply reduce to $\delta_{bc}$ as in~\eqref{eq:unit:def_unit}. If this is not the case, we can diagonalise the matrix, such that in the new basis (without Einstein-summation)
\begin{equation*}
 M_{ab} = w_a \delta_{ab}.
\end{equation*}
We are restricting ourselves to tensors that have a solution to~\eqref{eq:unit:gen_unit} such that all $w_a>0$. In the new basis, we can then redefine the tensor (without using Einstein-summation)
\begin{equation}\label{eq:unit:redef_P}
 \tilde{P}_{abc} = \frac{1}{\sqrt{w_a w_b w_c}} P_{abc},
\end{equation}
which means we get a new tensor $\tilde{P}_{abc}$ that has a unit given by $\mathbf{1} = \sum_a \tilde{\alpha}_a f_a = \sum_a \sqrt{w_a} \alpha_a f_a$. This can be seen by (without using Einstein-summation)
\begin{equation*}
 \sum_a \alpha_a \sqrt{w_a} \tilde{P}_{abc} = \sum_a \alpha_a P_{abc} \frac{1}{\sqrt{ w_b w_c}} = \frac{w_b \delta_{bc}}{\sqrt{w_b w_c} } = \delta_{bc}.
\end{equation*}

In order to understand under which conditions the matrix $M_{ab}$ is positive definite, consider the functional
\begin{equation}\label{eq:unit:functional_f}
 f(\alpha) := \alpha_a \alpha_b \alpha_c P_{abc},
\end{equation}
and suppose we are interested in finding the local extrema of this functional under the condition that $|\alpha|^2 = \alpha_a \alpha_a = 1$. This can be done by introducing a Lagrange-multiplier $k$,
\begin{equation*}
 g(\alpha) := \alpha_a \alpha_b \alpha_c P_{abc} + k \big(1-|\alpha|^2\big).
\end{equation*}
Finding the local extrema can then be done by taking the derivative with respect to $\alpha_a$,
\begin{equation*}
 \partial_{\alpha_c} g(\alpha) = 3\alpha_a \alpha_b P_{abc} - 2 k \alpha_c = 0.
\end{equation*}
Note that if we define $\alpha_a' = \frac{2k}{3} \alpha_a$, this exactly reproduces the eigen-problem of~\eqref{eq:unit:gen_unit}:
\begin{equation*}
 \alpha_a' \alpha_b' P_{abc} = \alpha_c'.
\end{equation*}
Thus, by the definition of the matrix $M_{ab}$, the extremal value $\alpha_a$ is an eigenvector of $M_{ab}$
\begin{equation*}
 \alpha_{a} M_{ab} = \alpha_a \alpha_c' P_{abc} = \frac{2k}{3} \alpha_a \alpha_c P_{abc} = \left(\frac{2k}{3}\right)^2 \alpha_b.
\end{equation*}
It can be seen that the eigenvalues are positive for real solutions. This also implies
\begin{equation*}
 \alpha_a \alpha_b \alpha_c P_{abc} = \frac{2k}{3}.
\end{equation*}
This sets the first eigenvalue to $\left(\frac{2k}{3}\right)^2$, for a condition for the other eigenvalues, let us consider a~second order perturbation of~\eqref{eq:unit:functional_f} denoted by $\epsilon \delta_a$, with $\delta_a$ a unit-size vector. It follows from the restriction $|\alpha|^2=1$ that $\alpha_a \delta_a = 0$. The first-order contribution is zero since we are considering the extremal value. The second-order contribution is given by
\begin{equation*}
 3\alpha_a \delta_b \delta_c P_{abc} = \frac{9}{2k} M_{bc} \delta_b \delta_c,
\end{equation*}
this means that the requirement that $M_{ab}$ is positive-definite coincides with the functional~\eqref{eq:unit:functional_f} either having a local minimum with a positive value ($k>0$), or a local maximum with a negative value ($k<0$). Here, we will choose the positive value as the canonical choice.

Let us remark on how to choose the $\alpha_a$ if there are several solutions to the above, such that the matrix $M_{ab}$ is positive definite. Since we are looking for ``almost unit'' functions, it makes sense to look at the minimal solution for{\samepage
\begin{equation*}
 \sum_{a=1}^N (w_a - 1)^2.
\end{equation*}
This would then produce a candidate unit which is as close to the real unit as possible.}

A potentially worrisome fact is that we are considering only tensors that have solutions such that $w_a>0$, in other words such that $M_{ab}$ is positive-definite. However, this might not be such a~big issue due to two main reasons. Firstly, in~\cite{Obster:2017pdq, Obster:2017dhx} it has been shown that the canonical tensor model wave functions tend to have peaks around symmetric configurations. This suggests that while quantum fluctuations from a symmetric spacetime might make $P_{abc}$ non-unital or non-associative, the deviation from this symmetric tensor is not expected to be that large, and thus the $w_a$ above will not deviate that much from the unit of the symmetric tensor. Secondly, if a tensor $P_{abc}$ does have negative eigenvalues, this might be a sign that this tensor does not correspond to a Riemannian manifold. Instead, it is expected that the tensor might describe a \emph{pseudo}-Riemannian manifold. As our current aim is to first develop an understanding of algebras corresponding to compact Riemannian manifolds, it is reasonable to leave these cases for future study.

The values $w_a$ have an interesting interpretation, that might be useful for model-building. Considering a compact Riemannian manifold $(\mathcal{M},g)$, we can equip the Hilbert space as introduced in Section~\ref{sec:topology} with a compact self-adjoint operator $\mathcal{O} \colon L^2(\mathcal{M}) \rightarrow L^2(\mathcal{M})$. According to the spectral theorem~\cite{conway1994course}, there exists an orthonormal basis of $L^2(\mathcal{M})$ consisting of eigenvectors of~$\mathcal{O}$. Let us now consider the basis $\{f_a \}$ we introduced in Section~\ref{sec:topology}, and assume that this basis consists of eigenvectors of $\mathcal{O}$. The eigenvalues $\lambda_a$, have the property $\lim_{a\rightarrow\infty} \lambda_a = 0$. Instead of $\tilde{P}_{abc} = \braket{f_c| f_a \cdot f_b}$, we could instead consider the tensor
\begin{equation}\label{eq:unit:op_new_tensor}
 P_{abc} := \braket{\mathcal{O} f_c | \mathcal{O}(f_a)\cdot \mathcal{O}(f_b)} = \lambda_a \lambda_b \lambda_c \tilde{P}_{abc}.
\end{equation}
Comparing this to~\eqref{eq:unit:redef_P}, this gives the $w_a$ an interpretation of the eigenvalues of a compact self-adjoint operator, according to the relationship
\begin{equation}\label{eq:unit:wa}
 w_a = \lambda_a^2.
\end{equation}
Here we will assume the eigenvalues of such an operator to be positive, such that the values $w_a$ can directly be interpreted without any loss of information.

In order to construct such a compact operator, let us consider a different operator called the Laplace--Beltrami operator. This is a prominent operator that has an intrinsic connection to the geometry of Riemannian manifolds, see Appendix~\ref{sec:app:laplace} for more information on the Laplace--Beltrami operator. This operator $\Delta \colon C^{\infty}(\mathcal{M}) \rightarrow C^{\infty}(\mathcal{M})$ is known to generate an orthonormal basis of $L^2(\mathcal{M})$ as described above, though it is not compact. Its eigenvalues are negative, so often we consider $-\Delta$ instead. The eigenvalues are denoted by $-\mu_a$ and given by
\begin{gather*}
 -\Delta f_a = \mu_a f_a,\qquad
 0 = \mu_1 < \mu_2 \leq \mu_3 \leq \mu_4 \leq \cdots ,
\end{gather*}
with its limit $\lim_{a\rightarrow\infty} \mu_a = \infty$. The intrinsic connection between this operator and the geometry (encoded by the metric) of a Riemannian manifold may already be seen in its representation in local coordinates
\begin{equation*}
 \frac{1}{\sqrt{\det g}} \partial_j \bigl(g^{ij} \sqrt{\det g} \partial_i f\bigr),
\end{equation*}
and in fact it is possible to actually determine the metric from the information of the Laplacian acting on the functions $f\in C^{\infty}(\mathcal{M})$~\cite{STRICHARTZ198348}. The Laplacian has been used in studies of the geometry in quantum gravity before, for instance in the case of causal dynamical triangulation~\cite{Ambjorn:2005db,Clemente:2018czn} and the canonical tensor model~\cite{Kawano:2018pip}.

Combining the information of the last two paragraphs, we find an interesting new interpretation of the $w_a$, and we have the opportunity to construct an operator $\mathcal{O}$ that not only serves to generate an orthogonal basis but also includes important geometric information. An example of such an operator would be
\begin{equation}\label{eq:unit:ex_operator}
 \mathcal{O} = \e^{\Delta}.
\end{equation}
A similar operator has actually been used in the canonical tensor model before to generate the tensor similarly to~\eqref{eq:unit:op_new_tensor}, with the reasoning that this would smoothen the cutoff when considering an $N$-dimensional tensor instead of an infinite-dimensional one~\cite{Kawano:2018pip}. In this work we find that the cutoff might actually not be that much of a problem, as we can reconstruct the topology and infinite-dimensional algebra using the associative closure. However, adding this extra operator actually adds geometric information which might in part explain the success of the approach in~\cite{Kawano:2018pip} to extract topological and geometric data using the tensor rank decomposition.

It still has to be seen which operator would be good to use in a certain context. It may be expected that for a certain model, for instance the canonical tensor model, one needs to specify which operator is considered in order to make the complete link to gravity.

\section{Examples: Riemannian manifolds and fuzzy spaces}\label{sec:cases}
In this section, we will discuss several examples of the formalism introduced in this work. The flat circle has been developed throughout this paper as a concrete example of how a tensor can correspond to a Riemannian manifold. We have shown that the topology, measure and the full algebra of functions can be reconstructed from the tensor constructed in Section~\ref{sec:topology}. It is remarkable that we can reconstruct this whole structure from a five-dimensional symmetric tensor with just $35$ entries. However, it is important to show that this formalism can handle a wide range of spaces. For this, we look at three main areas. We analyse the behaviour of perturbations of the flat circle in Section~\ref{sec:cases:pert}, then we analyse an inherently fuzzy space namely the semi-local circle in Section~\ref{sec:cases:semi-local}, and lastly we will consider the sphere in Section~\ref{sec:cases:sphere} to show that the framework works just as well in higher dimensions.

In this section, for every example a benchmark will be cited. This is the maximal error value~${\Delta > 0}$, below which a tensor rank decomposition will be accepted
\begin{equation*}
 \Bigg(P_{abc} - \sum_{i=1}^R \phi_a^i \phi_b^i \phi_c^i\Bigg)^2 < \Delta.
\end{equation*}
Often, this value can be taken to be extremely small \big($\Delta \sim O\big(10^{-30}\big)$\big), especially if the tensor has a certain continuous symmetry such that there are many minimal tensor rank decompositions, but in some cases where the minimal tensor rank decomposition is harder to find we have to increase this value. As a general rule, we will always require it to be at most $\Delta < 10^{-6}$.

\subsection{Perturbations around the flat circle}\label{sec:cases:pert}
Looking at perturbations around a given tensor is a useful first step to understanding the behaviour of space when altering the algebra of functions. Furthermore, given the strong peaks of the quantum wave function of the canonical tensor model around symmetric configurations~\cite{Obster:2017pdq}, it seems that we should expect small perturbations to occur around tensors representing smooth spaces.

As we have a relatively good understanding of the flat circle by now, let us consider perturbations around this tensor. The vector space $\mathcal{F}\cong \mathbb{R}^5$ is kept the same. We will denote the original tensor, as described in Section~\ref{sec:topology}, as $P_{abc}$ and any perturbed tensor by $\tilde{P}_{abc}$
\begin{equation*}
 \tilde{P}_{abc} = P_{abc} + Q_{abc},
\end{equation*}
where $Q_{abc}$ is the symmetric tensor characterising the perturbation. All tensors described here act on the same vector space $\mathcal{F}$, not on extensions of it or such. Generally, $\tilde{P}_{abc}$ will not be unital anymore, so we will have to generate a new unit according to the procedure in Section~\ref{sec:unit}. We will consider several different kinds of perturbations:
\begin{itemize}\itemsep=0pt
 \item Using the potential homomorphisms found using the tensor rank decomposition in Section~\ref{sec:associative_constr:pot_hom}, we perturb the $\beta_i$ measure factors.
 \item Using the potential homomorphisms found using the tensor rank decomposition in Section~\ref{sec:associative_constr:pot_hom}, we perturb one of the points $p^i$ themselves.
 \item We add a random ``high-energy'' perturbation, meaning a perturbation of $Q_{abc}$ where $Q_{abc}\neq 0$ if $a,b,c \geq 4$.
\end{itemize}
What we will see is that in the case of perturbing the tensor rank decomposition itself, it is reasonably well-behaved and the perturbation mainly results in a deformation of the manifold. When altering or adding points, it depends on the size of the perturbation. Small perturbations still remain smooth, whereas bigger perturbations or random high-energy perturbations lead to a breaking of the smooth structure and only a finite set of points remain.

It should be noted that for perturbations it was sometimes more difficult to find an exact minimal tensor rank decomposition. In principle, one could use a higher-rank decomposition as well, but then one has to filter out the decompositions that do not correspond to potential homomorphisms as explained in Section~\ref{sec:associative_constr:pot_hom}. In this work we take an approximate minimal tensor rank decomposition, where we are satisfied with an error $\big(P_{abc}-\sum_{i=1}^R \beta_i p^i_a p^i_b p^i_c\big)^2 < \Delta = 10^{-6}$. Since the main goal here is to demonstrate how these perturbations affect the shape of the functions, finding a method to find higher-rank tensor rank decompositions that correspond to potential homomorphisms only is left for later study. Furthermore, this section is not meant as a systematic analysis of perturbations using this framework but mainly a proof of concept to show the potential of types of spaces that can be analysed within this framework. There are a~lot of interesting things to find out about them, but this is out of the scope of the present paper.

\textbf{Case 1: Perturbing the measure.}
Perturbing the measure was done as follows. First the tensor $P_{abc}^{(u)}$ was defined as the five-dimensional tensor in Section~\ref{sec:topology}. Then, an extra damping factor was introduced using the operator $\mathcal{O} f_a = \e^{\Delta/25}f_a = \e^{-n_a^2/25}f_a$, with $n_a = \{0, 1, 1, 4, 4\}$, in order to demonstrate the reconstruction of the unital tensor $P_{abc}^{(u)}$ as described in Section~\ref{sec:unit},\footnote{Note that the factor $25$ is arbitrary, and any number could have been taken.}%
\begin{equation*}
 P_{abc} = \e^{-n_a^2/25 -n_b^2/25 -n_c^2/25} P_{abc}^{(u)}.
\end{equation*}
The unital tensor $P_{abc}^{(u)}$ was then reconstructed using the procedure in Section~\ref{sec:unit}. This means that the solutions to
\begin{equation*}
 \alpha_a \alpha_b P_{abc} = \alpha_c
\end{equation*}
were found. This was done using \textsc{Mathematica} and finding solutions to $\sum_{c=1}^5 (\alpha_a\alpha_b P_{abc} - \alpha_c)^2 = 0$ by using minimisation. Then, the matrix $M_{ab}$ was found (without Einstein-summation)
\begin{equation*}
 M_{ab} = \sum_{c=1}^5 \alpha_c P_{abc} = w_a \delta_{ab},
\end{equation*}
where $w_a$ are exactly given by
\begin{equation*}
 w_a = \big\{1, {\rm e}^{-2/25}, {\rm e}^{-2/25}, {\rm e}^{-8/25}, {\rm e}^{-8/25}\big\}.
\end{equation*}
Note that these exactly correspond to the square of the eigenvalues, as also expected from~\eqref{eq:unit:wa}. From this tensor,\footnote{Note that this tensor now includes some information about the metric. Here we will not go deeper into this, but it would be interesting to examine what part of the metric we can reconstruct from this more closely in a~future study.} we now reconstruct the unital tensor
\begin{equation*}
 P_{abc}^{(u)} = \frac{1}{\sqrt{w_a w_b w_c}} P_{abc}.
\end{equation*}

Using the tensor rank decomposition, as described in Section~\ref{sec:associative_constr:pot_hom}, $1400$ points $p^i$ with their measure-factor $\beta_i$ were generated. This gives a decomposition of
\begin{equation*}
 P_{abc}^{(u)} = \sum_{i=1}^{1400} \beta_i p_a^i p_b^i p_c^i.
\end{equation*}
This leads to an equivalent picture as in Figure~\ref{fig:ass_constr:circle_TRD}. In this case, we want to perturb the measure. The perturbation tensor is defined as
\begin{equation*}
 Q_{abc} = \epsilon \sum_i \beta_i' p_a^i p_b^i p_c^i,
\end{equation*}
where $\beta_i'$ characterises which points $p^i$ will be perturbed. We also added a factor of $\epsilon>0$ to alter the general size of the perturbation. This means that, for example, $\beta_i' = 0$ corresponds to no perturbation, or for example $\beta_1'=1$ corresponds to a perturbation of the first point. The measure factors $\beta_i$ were altered in two ways. The first approach was to change the measure factor for only the first point, $\beta_1$, such that
\begin{equation}\label{eq:cases:circ_beta:p1}
 \beta_i' = \delta_{i1}.
\end{equation}
The second approach was to alter all of the factors $\beta_i$, using a Gaussian distribution
\begin{equation}\label{eq:cases:circ_beta:ap}
 \beta_i' = \e^{-(i-700)^2/36}.
\end{equation}
The factor $36$ was taken such that there would be a reasonable range, such that not just a few points would be affected. The full new, perturbed, tensor is then given by
\begin{equation*}
 \tilde{P}_{abc} = P_{abc}^{(u)} + \epsilon Q_{abc} = \sum_{i=1}^{1400} (\beta_i+\epsilon\beta_i') p_a^i p_b^i p_c^i,
\end{equation*}
hence the claim that we are perturbing the measure.

\begin{figure}[ht]
 \centering
 \begin{minipage}{0.4\textwidth}
 \includegraphics[width=0.99\textwidth]{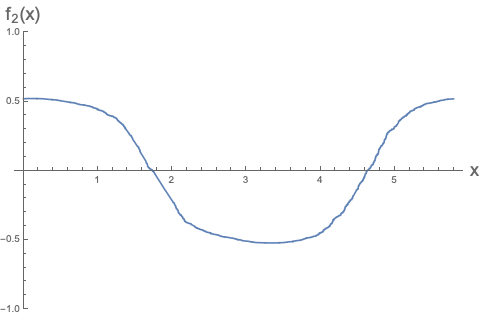}
 \end{minipage}
 \begin{minipage}{0.4\textwidth}
 \includegraphics[width=0.99\textwidth]{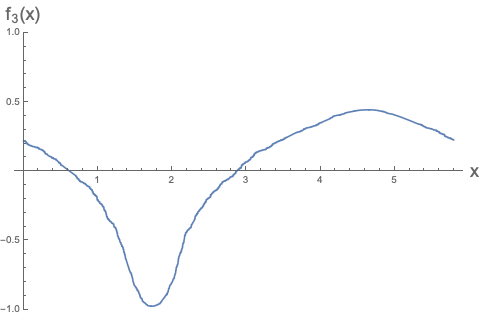}
 \end{minipage}
 \begin{minipage}{0.4\textwidth}
 \includegraphics[width=0.99\textwidth]{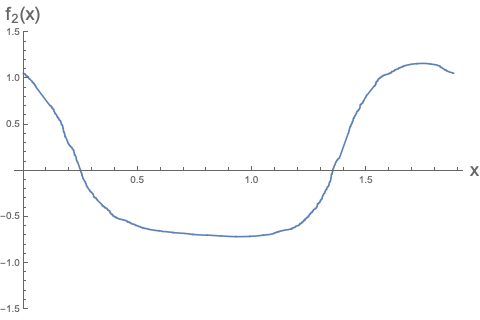}
 \end{minipage}
 \begin{minipage}{0.4\textwidth}
 \includegraphics[width=0.99\textwidth]{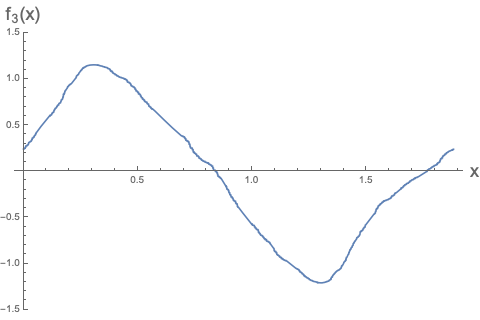}
 \end{minipage}

 \caption{A plot of the functions $f_2(x)$ and $f_3(x)$ of the circle with a perturbed measure, where the function values are defined as in~\eqref{eq:associative_closure:measure:f_representation}. The top two images represent the case of~\eqref{eq:cases:circ_beta:p1} using $\epsilon=0.4$ and $\Delta=7.6\cdot 10^{-31}$. The bottom two images represent the case of~\eqref{eq:cases:circ_beta:ap} using $\epsilon=0.5$ and $\Delta=2.8\cdot 10^{-30}$. In both cases, 200 tensor rank decompositions of rank 7 were used, yielding 1400 points.}
 \label{fig:cases:circle_beta}
\end{figure}

After altering the $\beta_i$'s, a new unit has to be found according to the procedure in Section~\ref{sec:unit}, since generally the new tensor $\tilde{P}_{abc}$ is not unital anymore. Then, the tensor rank decomposition method of Section~\ref{sec:associative_constr:pot_hom} was applied to find the potential homomorphisms, and they were ordered in the same way as for the flat circle in~\eqref{eq:duality:example:topology_norm}. The tensor rank used remained unchanged at $R=7$, and $1400$ points were generated this way. The results of this are displayed in Figure~\ref{fig:cases:circle_beta}.

Three things are interesting here. Firstly, the measure is clearly disturbed which can be seen by the deformation of the functions and the size of the circle (in both cases it became smaller), but the general properties of the circle are still present. This is encouraging, since it means that we can describe interesting spaces that do not exactly correspond to the algebra of a circle. Secondly, the reason why we can find this continuous amount of potential homomorphisms, is because the associative extension was defined in terms of any maximal partial algebra. In the case of the flat circle, there was a partial algebra $\mathcal{S}_3$ that already contained all possible potential homomorphisms, but very few potential homomorphisms in Figure~\ref{fig:cases:circle_beta} actually correspond to homomorphisms of the same partial algebra. Since we allow them to come from different maximal partial algebras we get this rich structure. It should be noted that all of these points in the associative closure will become proper homomorphisms. Lastly, due to the defining of a new unit, there is a sort of mixing that happens. Changing $\beta_1$ only does not simply make that part of the dual space bigger, but deforms the region around it too. Note that the perturbations here are by no means small, so the strong deformations are to be expected.

\begin{figure}[t]
 \centering
 \begin{minipage}{0.315\textwidth}
 \includegraphics[width=0.99\textwidth]{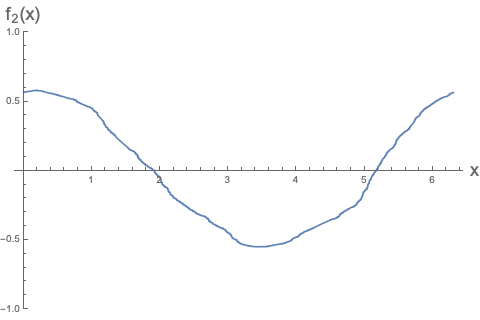}
 \end{minipage}
 \begin{minipage}{0.315\textwidth}
 \includegraphics[width=0.99\textwidth]{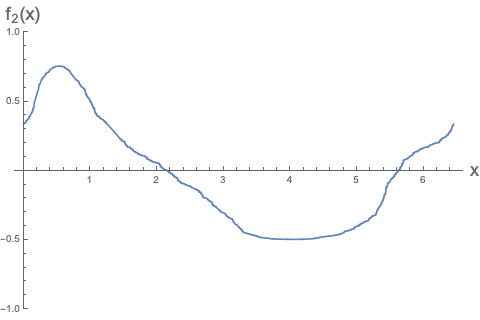}
 \end{minipage}
 \begin{minipage}{0.315\textwidth}
 \includegraphics[width=0.99\textwidth]{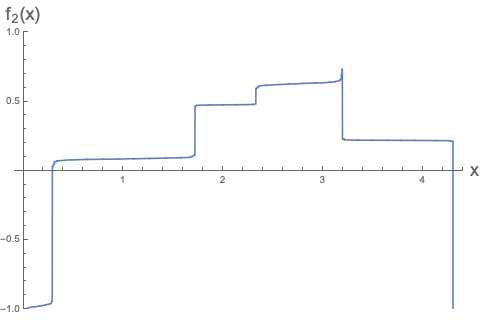}
 \end{minipage}

 \caption{A plot of the functions $f_2(x)$ of the circle with a shifted point as described in~\eqref{eq:cases:circ_shift}, where the function values are defined as in~\eqref{eq:associative_closure:measure:f_representation}. On the left using $\epsilon=0.1$, $\Delta=8.1\cdot 10^{-31}$, in the middle $\epsilon=0.25$, $\Delta=5.6\cdot 10^{-30}$ and on the right $\epsilon=0.75$, $\Delta=2.3\cdot 10^{-28}$. In all cases, 200 tensor rank decompositions of rank 7 were used, yielding 1400 points with the random perturbation vector of~\eqref{eq:cases:circle_pert_vector}, and the displayed functions are in the original basis in order to make direct comparison to the smooth functions on a circle possible. Note that in the plot on the right, many of the points overlap.}
 \label{fig:cases:circle_shift}
\end{figure}

\textbf{Case 2: Perturbing a point.}
The perturbation of a point was done by shifting one point of a tensor rank decomposition. This was done using the unperturbed tensor rank decomposition $P_{abc}=\sum_{i=1}^{7} \beta_i p_a^i p_b^i p_c^i$ with 7 points as discussed above. The first point $p^1$, which should be noted is arbitrarily picked, was taken and shifted with a vector $q$
\begin{equation*}
 p^1_a \rightarrow p^1_a + \epsilon q_a,
\end{equation*}
where again $\epsilon>0$ is introduced to set the size of the perturbation. This yields a new tensor with a tensor rank decomposition
\begin{equation*}
 \tilde{P}_{abc} = \beta_1 \big(p^1 + \epsilon q\big)_a \big(p^1 + \epsilon q\big)_b \big(p^1 + \epsilon q\big)_c + \sum_{i=2}^{7} \beta_i p_a^i p_b^i p_c^i .
\end{equation*}
In terms of the perturbation tensor $Q_{abc}$, this corresponds to
\begin{equation}\label{eq:cases:circ_shift}
 Q_{abc} = \epsilon \beta_1 \big(p^1_a p^1_b q_c + p^1_a q_b p^1_c + q_a p^1_b p^1_c + \epsilon p^1_a q_b q_c + \epsilon q_a p^1_b q_c + \epsilon q_a q_b p^1_c + \epsilon^2 q_a q_b q_c \big).
\end{equation}
A similar procedure as in the previous case was then followed, taking a random normalised vector
\begin{equation}\label{eq:cases:circle_pert_vector}
 q=(0, 0.714853, -0.514041, -0.247008, 0.40464),
\end{equation}
generated by \textsc{Mathematica}. Note that in order to not interfere too much with the unit, the first component of the vector was taken to be $0$. Then, the algebra was redefined as in Section~\ref{sec:unit} and tensor rank decompositions were generated just like above. The result of this can be found in~Figure~\ref{fig:cases:circle_shift}.

Interestingly, for small values of $\epsilon$, the functions do not change that much and the shape of the functions is just deformed, similarly to changing the measure as described above. However, when the perturbation becomes large, $\epsilon \sim O(1)$, the smooth structure breaks down.

\textbf{Case 3: Random high-energy perturbation.}
Adding a random perturbation is a very uncontrolled way of adding a perturbation, but nonetheless it is interesting because a priori it is not known what kind of perturbations one should expect from a quantum theory. The perturbation is now characterised by the tensor
\begin{equation}\label{eq:cases:circle_HE}
 Q_{abc} = \begin{cases} \epsilon \operatorname{RandomReal}(-1,1)& \text{if}\ a,b,c \geq 4,\\
 0 & \text{otherwise}.
 \end{cases}
\end{equation}
Here $\operatorname{RandomReal}$ denotes a function giving a random number in the range $[-1,1]$ according to a uniform distribution. For $\epsilon$, a value of $\epsilon=0.01$ was chosen. We can characterise this perturbation by four elements; $Q_{444}$, $Q_{445}$, $Q_{455}$, and $Q_{555}$. In the current example, we used specifically (randomly generated)
\begin{gather}
 Q_{444} = -0.00616405,\qquad
 Q_{445} = 0.000446477,\qquad
 Q_{455} = -0.00686687,\nonumber\\
 Q_{555} = -0.00208394.\label{eq:cases:circle_pert:Q_random}
\end{gather}
Then, the same procedure was repeated as before. Note firstly that the tensor $Q_{abc}$ does not alter the unit of the tensor, since $Q_{1bc} = 0$, so finding a new unit is not necessary. Furthermore, in this case it appears easier to find a higher rank-decomposition and required its to lie in the dual space of the partial algebra generated by $\mathcal{S}_3=\{f_1, f_2, f_3\}$, since this is still a covering partial algebra.\footnote{Note that in a future study it would be interesting to include other partial algebras into this picture too, as they might not all be equivalent anymore due to this perturbation.} Adding the perturbation will break some of the symmetric properties, so it is expected that this tensor will not correspond to a smooth space anymore.

\begin{figure}[t]
 \centering
 \begin{minipage}{0.31\textwidth}
 \includegraphics[width=0.99\textwidth]{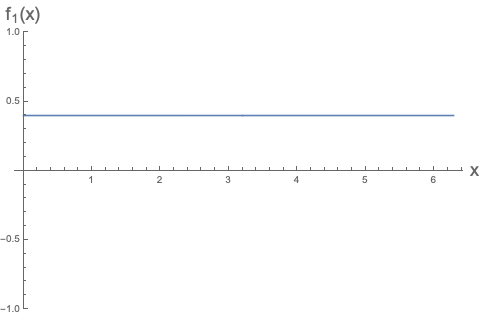}
 \end{minipage}
 \begin{minipage}{0.31\textwidth}
 \includegraphics[width=0.99\textwidth]{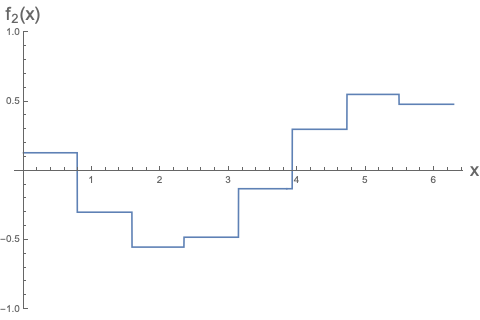}
 \end{minipage}
 \begin{minipage}{0.31\textwidth}
 \includegraphics[width=0.99\textwidth]{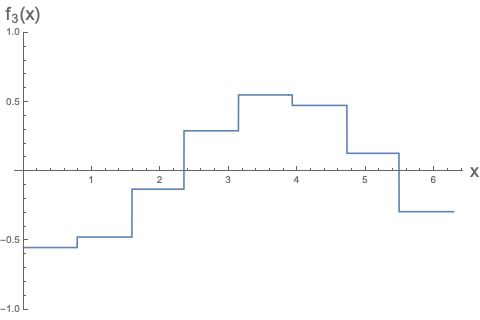}
 \end{minipage}

 \caption{A plot of the functions $f_1(x)$, $f_2(x)$ and $f_3(x)$ of the circle with a random high-energy perturbation as described in~\eqref{eq:cases:circle_HE}, where the function values are defined as in~\eqref{eq:associative_closure:measure:f_representation}, using the perturbation tensor as~\eqref{eq:cases:circle_pert:Q_random}, and generating 87 rank-8 tensor rank decompositions yielding 696 points (though many of them overlap).}
 \label{fig:cases:circle_HE}
\end{figure}

The result of this procedure is given in Figure~\ref{fig:cases:circle_HE}. It can be seen that apparently, for this perturbation, the resulting tensor has $8$ potential homomorphisms. It is expected that for higher energy perturbations (meaning perturbations for $a,b,c>M$ for some large $M$) there would be many more potential homomorphisms but still a possible breaking of the smooth structure. This suggests that, if the quantum theory for gravity admits any random perturbations, spacetime on a trans-Planckian level is actually discrete, though it could also be the case that symmetric configurations with smooth emerging spaces might be preferred as suggested in the canonical tensor model~\cite{Obster:2017pdq, Obster:2017dhx}.

A last note here to the general structure of the definitions in Section~\ref{sec:associative}. The definitions of the associative extension and associative closure attempt to make sure that as much as possible of the partial algebras is retained. It might, in the future, turn out that it is more beneficial to relax the requirements slightly to allow more points. At present, it is not clear what would be a better way to define it, so in this picture these kind of perturbations likely lead to some fundamental discreteness of space.

\subsection{The semi-local circle}\label{sec:cases:semi-local}
The semi-local circle is an example of a tensor that is expected to reproduce a fuzzy space. It is defined using a constant $0<\alpha<1$ and $a,b,c \leq N$ for some positive integer $N$ as
\begin{equation}\label{eq:cases:semi_local_P}
 P^\alpha_{abc}=\begin{cases}
 1 & \text{if}\ a=b=c,\\
 \alpha & \text{if}\ a=b=c\pm 1 \parallel a=b\pm 1=c \parallel a\pm 1 = b = c,\\
 0 & \text{otherwise},
 \end{cases}
\end{equation}
where we use circular boundary conditions, i.e., $N+1 = 1$ and $1-1 = N$. It has a discrete version of a translation symmetry, as it is symmetric under $(a,b,c) \rightarrow (a+1,b+1,c+1)$. We consider a set of local functions as having the property that (without Einstein-summation)
\begin{equation*}
 f_a \cdot f_b = \delta_{ab} f_a,
\end{equation*}
which would be the case for
\begin{equation*}
 P_{abc} = \begin{cases}
 1 & \text{if } a=b=c,\\
 0 & \text{otherwise}.
 \end{cases}
\end{equation*}
Together with the cyclic property it possesses this explains the terminology ``semi-local circle''.

\begin{figure}[t]
 \centering
 \begin{minipage}{0.41\textwidth}
 \includegraphics[width=0.99\textwidth]{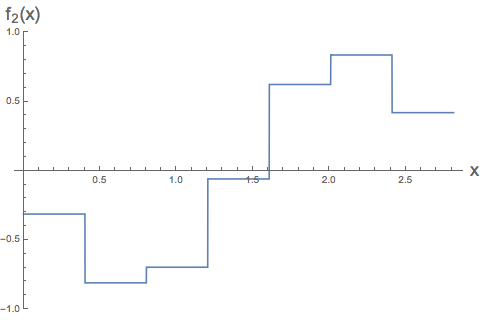}
 \end{minipage}
 \begin{minipage}{0.41\textwidth}
 \includegraphics[width=0.99\textwidth]{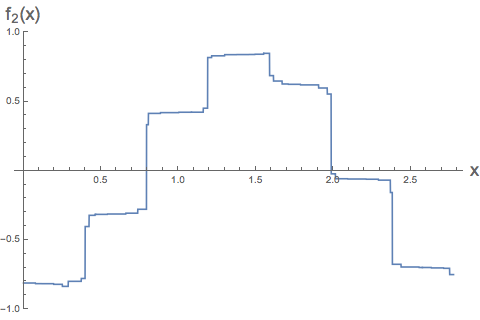}
 \end{minipage}
 \begin{minipage}{0.41\textwidth}
 \includegraphics[width=0.99\textwidth]{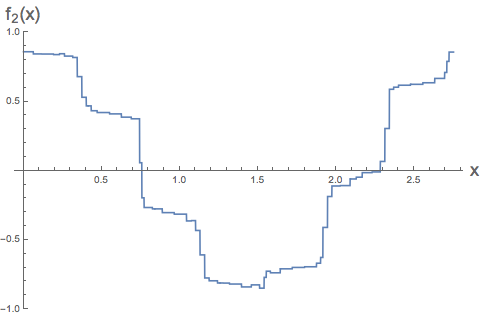}
 \end{minipage}
 \begin{minipage}{0.41\textwidth}
 \includegraphics[width=0.99\textwidth]{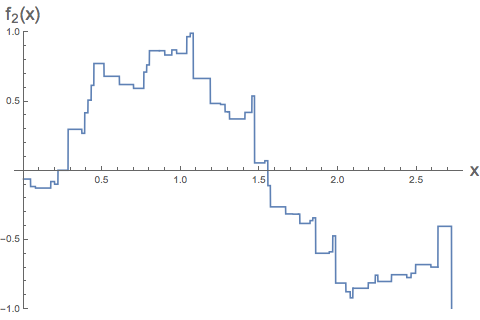}
 \end{minipage}

 \caption{A plot of the functions $f_2(x)$ of the semi-local circle with $\alpha=0.1$ as defined in~\eqref{eq:cases:semi_local_P}, where the function values are defined as in~\eqref{eq:associative_closure:measure:f_representation}, generated using various ranks of tensor rank decompositions. From top-left to bottom-right, they are $R=7, 9, 10, 11$. The amount of points generated are, respectively,~$70$,~$90$,~$100$,~$99$. Note that the first point is always selected at random, so there is no meaning behind the phase-difference.}
 \label{fig:cases:semi-circle}
\end{figure}

This tensor gives an interesting example because of its discrete nature and easy way to construct. Besides, since it is not constructed using a smooth background like the examples in Section~\ref{sec:cases:pert}, it is an example of an intrinsically algebraic space.

The procedure followed to analyse this tensor is the same as before. First one generates a~unit and redefines the tensor as in Section~\ref{sec:unit}, then one uses the tensor rank decomposition of Section~\ref{sec:associative_constr:pot_hom} to generate potential homomorphisms. This was done for $N=7$. As mentioned before, we always target a $\Delta<10^{-6}$, however in this case we also evaluated decompositions with a lower rank, as there is some interesting behaviour going on. The results for $f_2(x)$ for $R=7, 9, 10, 11$ are shown in Figure~\ref{fig:cases:semi-circle}.

Interestingly, though the $R=7$ case is not very precise with a $\Delta=10^{-2}$, but it shows a~discretised space of $7$ points, similar to what one would naively expect from the tensor described above. However, since the decomposition does not seem to be correct, one should increase $R$. For $R=9$ the decomposition is still not great with $\Delta=4\cdot 10^{-3}$, but some actual fuzzy behaviour can be observed which only increases with $R=10$. Finally, with $R=11$ proper decompositions can be found with $\Delta = 10^{-25}$. Interestingly we see that it starts to look more like a circle, but with some fuzzy points included and the regions that were present in the smaller $R$ cases dominate with large steps in between. It seems to be a very interesting example that would be insightful to investigate further, for instance by taking various different $N$ and checking if all points indeed are included in the dual space of some system of partial algebras. In the current check this has not been performed rigorously, since one needs to find a candidate maximal system of algebras first which is more difficult than in the case of the circle or the sphere. This would be interesting to take a look at in a future study.

\subsection{The sphere}\label{sec:cases:sphere}
The sphere is an example of a 2-dimensional Riemannian manifold. This example is taken to demonstrate the ability of this framework to handle higher-dimensional manifolds as well.
The tensor representing the sphere is found using the procedure in Section~\ref{sec:topology}. First, one chooses a~basis for the real smooth functions $C^\infty\big(S^2\big)$. One such basis is given by the real spherical harmonics, $Y_l^m(\theta,\phi)$, which are eigenfunctions of the Laplace--Beltrami operator. The ordering of eigenfunctions is determined by $a\in\{1,2,3,\dots\} \rightarrow (l,m) \in \{(0,0), (1,-1), (1,0), (1,1), (2,-2),\dots \}$. The tensor $P_{abc}$ is then generated by
\begin{equation*}
 P_{abc} = \int_{S^2} {\rm d}\Omega \, Y_a(\Omega) Y_b(\Omega) Y_c(\Omega),
\end{equation*}
where $\Omega=(\theta,\phi)$ and ${\rm d}\Omega = \sin(\theta) {\rm d}\theta {\rm d}\phi$.

In this example, the smallest non-trivial tensor was used corresponding to the case where we cut the tensor off at $N=9$, such that the products for $l=\{0,1\}$ are all fully included. Determining the potential homomorphisms can be done in two ways, either by evaluating a~similar equation as in~\eqref{eq:duality:example_homo}
\begin{equation*}
 p\in\mathbb{R}^9,\ a,b \in \{1,\dots,4\} \colon \ p_a p_b = \sum_{c=1}^9 P_{abc}p_c,
\end{equation*}
or by using the tensor rank decomposition. Here, the former approach was used to construct $1200$ points.

A more difficult thing is how to illustrate the topology of the space. In Section~\ref{sec:topology}, the proper mathematical way of defining the topology was described in Figure~\ref{fig:duality:topology}. Even though this definition is unambiguous, in practice it is more difficult to visualise this in a two-dimensional case. Here a simple way was considered, since this is only meant to be a proof of concept, but in the future it would be good to have a more rigid approach.

\begin{figure}
 \centering
 \begin{minipage}{0.42\textwidth}
 \includegraphics[width=0.95\textwidth]{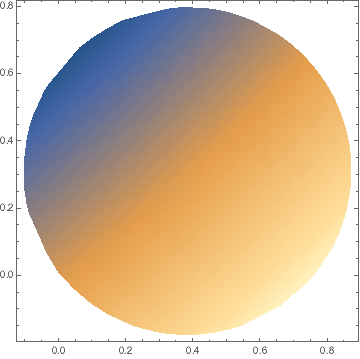}
 \end{minipage}\qquad
 \begin{minipage}{0.42\textwidth}
 \includegraphics[width=0.95\textwidth]{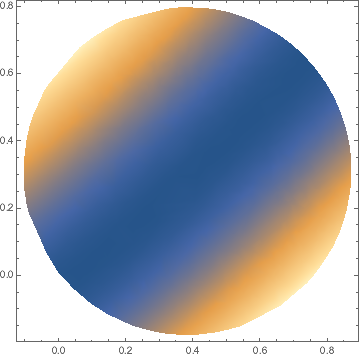}
 \end{minipage}

 \caption{Density plots of the functions $f_3(x,y)$ on the left, supposed to represent $Y^1_0$, and $f_7(x,y)$ on the right, supposed to represent $Y^2_0$, using the definition of the coordinates as defined in the text. Though the way of defining the coordinates in the text is not perfect, a clear resemblance to the spherical harmonics may be observed. 1200 points were used to create these density plots.}
 \label{fig:cases:spherical}
\end{figure}

One way that can be used is by defining a topological distance between the points as
\begin{equation*}
 d\big(p^i, p^j\big) = \sqrt{\sum_{a=1}^N |p_a^i - p_a^j|^2},
\end{equation*}
similar to the approach for the circle. Taking as a reference point $p^1$, we can then choose two points close to $p^1$, let's say $p^q$ and $p^r$. Then we define the direction $p^q$ as the $x$-axis, and the length of $p^q$ is simply $d\big(p^1, p^q\big)$. The $y$-axis is then found by solving
\[
d\big(p^1, p^r\big) = \sqrt{(p^r_x)^2 + (p^r_y)^2} \qquad \text{and} \qquad d\big(p^q, p^r\big) = \sqrt{(p^q-p^r_x)^2 + (p^r_y)^2}.
\]
Now we have defined an $x$- and $y$-axis and we can define a position of every point $p^i$ in the $(x,y)$-plane by solving
\[
d\big(p^q,p^i\big) = \sqrt{\big(p^q - p^i_x\big)^2 + \big(p^i_y\big)^2}\qquad \text{and}\qquad d\big(p^r, p^i\big) = \sqrt{\big(p^r_x - p^i_x\big)^2 + \big(p^r_y - p^i_y\big)^2}.
\]
 Note that this is a very crude way of defining part of the topology, as we use a local coordinate system and naively extrapolate this to the whole space. In Figure~\ref{fig:cases:spherical}, the result of this is shown for two of the eigenfunctions, however it should be noted that using this way of defining coordinates, not all eigenfunctions generally look this nice as for some eigenfunctions the ``front'' and ``back'' values of the sphere are mixed beyond a small local patch.

The points in Figure~\ref{fig:cases:spherical} were constructed using the dual space of the partial algebra, an approach already presented in Section~\ref{sec:topology} in the example of the flat circle. No information on the measure or geometry has been taken into account, as this is merely meant as a proof of concept to describe higher-dimensional spaces using this algebraic interpretation. However, it could have also been done using the tensor rank decomposition. A sanity check was performed, using $\Delta = 5\cdot10^{-12}$, it is possible to find rank $R=18$ tensor rank decompositions. The measure will also be regained when doing this, showing that the volume of the sphere is indeed $\sum_{i=1}^R \beta_i = 4\pi$ (up to 6 decimal places). More research has to be done here, to find a good way to visualise these properties properly and show exactly what the measure looks like.

\section{Implications for the canonical tensor model}\label{sec:CTM}
The main goal of this work was to lay a foundation for the understanding of the emergence of spacetime in the canonical tensor model (CTM). While it is not fully clear if this is the best and only spacetime interpretation of the CTM, at the very least this seems like an interesting approach to build a tensor-model that can describe (fuzzy) spaces, in a very different way than the original tensor models~\cite{Ambjorn:1990ge,Godfrey:1990dt,Sasakura:1990fs}. In this section, we will highlight some of the benefits of this interpretation for such a model, by interpreting important previous results of both the links of the CTM to gravity and the quantum version of the model. For a brief introduction to the CTM, we would like to refer to Appendix~\ref{sec:app:CTM}.

The CTM is a tensor model where the fundamental configuration is given by a real symmetric tensor of order three, $P_{abc}$. It is built in the canonical framework in order to introduce time to a~tensor model. This means in the present interpretation that the CTM corresponds to an algebra of functions over a (fuzzy) space, and both the topology through Section~\ref{sec:associative} and the geometry through Section~\ref{sec:unit} are allowed to fluctuate. Evaluating the time-evolution would then correspond to adding new spatial slices, in a similar sense to the ADM-formalism~\cite{Arnowitt1959}.

One important immediate result from the formalism presented here, is that in order to analyse countably infinite-dimensional function spaces it is possible to analyse the finite-dimensional counterpart and assume the associative closure. This makes analysis of a finite-dimensional tensor useful in the first place, as in~\cite{Sasakura:2006pq} it was already pointed out that finite-dimensional tensors could only directly correspond to completely local associative algebras. With the formalism developed here, we have shown that we could for instance take only the $5$-dimensional tensor written in Section~\ref{sec:topology} and reconstruct the full topology and measure on the circle. Besides, while to keep the mathematical formulation more simple we often assumed a finite-dimensional tensor, but the formalism seems to be able to be extended to countably infinite-dimensional tensors. This means that whether or not the final theory of quantum gravity is actually infinite-dimensional or finite-dimensional, the formalism seems to be able to handle both cases as long as the space is assumed to be compact.

In prior work, the authors already found that one can extract a remarkable amount of topological and geometric information from tensors~\cite{Kawano:2018pip, Kawano:2021vqc}, using tools from data analysis. Especially the tensor rank decomposition and persistent homology turned out to be useful. In the present work we discover why these tools actually work so well. Firstly, the tensor rank decomposition indeed generates points of the dual space of an algebra according to Section~\ref{sec:associative_constr:pot_hom}, which is the reason why using the tensor rank decomposition worked so well. Furthermore, if we choose a~discrete number of these points we argued in~\eqref{eq:duality:example:topology_norm} that two points are to be considered close in a topological sense if their inner product in $\mathbb{R}^N$ is large. This is exactly what was used in~\cite{Kawano:2018pip} in order to use persistent homology to extract topological data. In these papers, a damping function was introduced corresponding to~\eqref{eq:unit:ex_operator}. The original idea was to use this damping in order to make the cutoff more smooth and it was found that the results for extracting the topology and geometry became better. In the current paper we find that the inclusion of these damping functions actually corresponds to including explicit geometric information in the tensor. It appears crucial to reconstruct the full Riemannian manifold.

In~\cite{Sasakura:2014gia}, one of the first explicit connections between the CTM and general relativity was made. Here the authors investigated the $N=1$ case, and found that when rewriting the Hamiltonian to an action, the action exactly corresponded to the minisuperspace action of the Friedmann--Robertson--Walker (FRW) universe. This result fits the current framework very well, as here the $N=1$ case would mean that we are only considering the constant function over a~space. In other words, the $N=1$ case is the case of a completely isotropic universe, and the fact that the CTM leads to the FRW universe is an important check for the consistency of the CTM and specifically when interpreted using the framework developed in this work.

In terms of the quantum CTM, there has been a lot of work on the finding and analysing of wave functions~\cite{Kawano:2021vqc,Lionni:2019rty,Narain:2014cya,Obster:2017pdq,Obster:2017dhx,Obster:2020vfo,Sasakura:2021lub,Sasakura:2019hql}. Here it was shown that the CTM seems to prefer symmetric tensors, since there is evidence for strong peaks of these in a prominent wave function~\cite{Obster:2017pdq, Obster:2017dhx}. In the current context this would imply that the CTM, as already speculated, seems to have good indications for the emergence of macroscopic spaces. For instance a smooth manifold with a lot of symmetry seems to be preferred over a very odd-behaved fuzzy space.

There are also two main papers considering a formal continuum limit of the CTM~\cite{Chen:2016ate, Sasakura:2015pxa}. The formalism developed in the present work is not directly applicable to these cases, since these cases consider an emergent space~$\mathbb{R}^d$ which is not compact and thus the smooth functions~$C^\infty(\mathcal{M})$ do not have a Schauder basis which is countable. However, the insights from these papers might be important in the future to extend the formalism to non-compact spaces.

\section{Summary and future prospects}\label{sec:summary}
Finding a consistent theory of quantum gravity starts with a seemingly simple question: ``How do we describe gravity?''. In this work, we introduce a new way to describe a gravitational theory in terms of tensors. While tensor models, such as the canonical tensor model~(CTM)~\cite{Sasakura:2011sq}, already existed, it is important to fully understand how one can translate their results to a~spacetime picture as we expect from a theory of quantum gravity. The benefit of the framework introduced is that it is possible to describe all topological and geometric information of a compact Riemannian manifold in an object that only has countably infinite-dimensional degrees of freedom. Besides, it makes sense of finite-dimensional tensors too.

In this work we introduced the concept of an associative closure in order to deal with tensors that do not generate associative algebras. This helps to make the approach more stable, as was seen, for instance, in Section~\ref{sec:cases:pert} where the flat circle was perturbed. Furthermore, this made sure that even finite-dimensional tensors can correspond to smooth manifolds, making finite-dimensional calculations worthwhile. As it turned out, the symmetry of a tensor in the first two indices corresponds to generating a commutative algebra, but requiring the tensor to be fully symmetric and supposing a Hilbert-space structure gives us the opportunity to define a~measure. It was then argued that it is possible to introduce geometric information by means of the Laplacian or a similar operator into this formalism.

In principle, it is thus possible to have a countably infinite-dimensional symmetric order-three tensor $P_{abc}$ describing the full geometry of a compact Riemannian manifold. If one were only interested in the topology and measure, even a finite-dimensional tensor would suffice. This means that this would present a potential interpretation for models like the CTM, since except for a formal continuum limit this model generally works with finite-dimensional tensors. In Section~\ref{sec:CTM}, several implications for the CTM were discussed, and it was found that it seems consistent with several results. Especially since the CTM seems to prefer symmetric tensors, this interpretation would yield strong evidence for the emergence of macroscopic spacetimes.

There is a lot of interesting areas with research opportunities with this kind of framework. They are discussed below in three categories: The framework itself, the connection to the CTM, and the opportunity to build tensor models.

Firstly, the framework itself still has several places where it can be refined. For instance, in Section~\ref{sec:associative_constr:pot_hom}, two conjectures were phrased, which would be interesting to see if there is a way to prove these right or wrong, and under what conditions. Furthermore, there are still several questions in relation to Section~\ref{sec:unit}. Though for an infinite-dimensional tensor we can reconstruct the full geometry if we know the deviation of the candidate for a unit as generated by the tensor~$P_{abc}$ with the actual unit, for a finite-dimensional tensor this interpretation is less clear. One should either extend the operator to the full associative closure somehow, or only define the metric using functions up to the dimension of the tensor $N$ which will only give an approximate metric (which could be interpreted as fuzziness). Another question is about the case where the~$w_a$ as in~\eqref{eq:unit:redef_P} are not strictly positive. It is expected that we can still make sense of this as a~manifold, possibly as a pseudo-Riemannian manifold. Besides, we demonstrated in this paper how to reconstruct the topology and measure of a space, but it would be good to demonstrate that one could explicitly reconstruct the metric. A last remark about the framework itself goes to the example spaces that have been considered. While the analysis has been done thoroughly, the main goal was to provide ``proof of concepts'', but it would be very interesting to analyse these examples further and understand more of the properties of these emergent (fuzzy) spaces. It would be interesting to dive deeper into the work developed in~\cite{Barrett:2019aig, Glaser:2019lck, Glaser:2019lcd} as well, and apply learnings to the current work.

A second interesting research area relates to the CTM. We already discussed some of the implications for this model in Section~\ref{sec:CTM}, but more has yet to be done. For instance, several extensions~\cite{Narain:2015owa, Narain:2016sqn} and use cases~\cite{Sasakura:2014yoa, Sasakura:2014zwa, Sasakura:2015xxa} of the CTM have been identified in the past and it would be interesting to revisit them and see if there is any interesting interpretation of those. Furthermore, this paper is mainly looking at the kinematical question ``what is the interpretation of the object are we dealing with?'', but it would be very interesting to see if we can use the dynamics of the CTM to see how the emergent spaces would behave through time. Related to the last remark, it would be interesting to see if the exact spaces introduced here, or exact spaces with a modified unit giving them an interesting geometric interpretation according to Section~\ref{sec:unit}, are actually included in the tensors highlighted by the CTM wave-function~\cite{Obster:2017pdq, Obster:2017dhx}. This would be strong evidence for the quantum-CTM to reproduce macroscopic spaces.\looseness=1

A last research area to consider is the construction of other models using this framework. While the CTM seems to be a good candidate to apply this framework to, it is not sure if that is actually the case. It might be that this framework is best-used for a different kind of model. From Section~\ref{sec:unit} we know that one could attempt to construct a model with any operator that satisfies certain requirements. It would be interesting if we could derive a Hamiltonian for instance, starting with a certain operator and seeing how these equations should be to yield a~theory similar to general relativity. And in the case of the CTM: Which operator should we use to relate it to general relativity? Lastly, how can we describe other matter- and force-fields in this description. While they can be described as a manifold (fibre bundles) and are thus subject to the same kind of algebras as all other manifolds, they are expected to behave very differently. It might even be possible to give some meaning to non-commutative cases in this context, since from a tensor-model perspective this simply means dropping the symmetry-requirement on the tensor.

The connection between tensors, algebras and their dual topological spaces seems to open up a different way of interpreting gravity. The duality between geometry and algebra has been an active field of mathematics for a while, and combined with tensor models, it seems to provide an opportunity to describe spacetime in a way that would lead to a mathematically more straightforward formulation of quantum gravity.

\newpage

\appendix
\section{Measure theory}\label{sec:app:measure}
Throughout this work, some notions from measure theory are used and discussed. This section serves to fix some of the definitions. For a comprehensive introduction to the topic, we would like to refer to the literature, for instance~\cite{cohn2013measure}.

The general idea behind measure theory is to define a general notion of ``volume of a subset'', and use this to define a way of integration that is more generally applicable than Riemann integrals. The volume of a subset is basically a function which takes a subset and generates a positive real number, called a \emph{measure}. However, it is generally not possible to consistently define a measure that works on any collection of subsets, so one first needs to define what subsets one considers. This is done by means of a $\sigma$-algebra.
\begin{Definition}
 Let $X$ be an arbitrary set. A collection $\mathcal{A}$ of subsets is called a $\sigma$-algebra if%
 \begin{itemize}\itemsep=0pt
 \item $X\in \mathcal{A}$,
 \item for each set $A\in \mathcal{A}$, $A^c \in \mathcal{A}$,
 \item for each infinite sequence $\{A_i\}$ with $A_i \in \mathcal{A}$, the set $\bigcup_i A_i \in \mathcal{A}$,
 \item for each infinite sequence $\{A_i\}$ with $A_i \in \mathcal{A}$, the set $\bigcap_i A_i \in \mathcal{A}$.
 \end{itemize}
\end{Definition}
A $\sigma$-algebra $\mathcal{A}$ is basically a family of subsets of a set $X$ on which we can consistently define a measure. For this reason, a subset $A\in\mathcal{A}$ is called a \emph{measurable set}. An important $\sigma$-algebra that is mentioned in the text is the \emph{Borel $\sigma$-algebra} of a topological space $\mathcal{T}$, denoted $\mathcal{B}(\mathcal{T})$. It is generated by the open sets in a topological space, in the sense that one takes all open sets, and takes infinite intersections of these to add them to the Borel $\sigma$-algebra. This means that there is a canonical way to define a $\sigma$-algebra from a topological space.

We are now set to define a measure.
\begin{Definition}
 A \emph{measure} is a function from a $\sigma$-algebra $\mathcal{A}$ to the positive real numbers
 \begin{equation*}
 \mu \colon\ \mathcal{A} \rightarrow \mathbb{R}_+ \cup \{+\infty\},
 \end{equation*}
 that is countably additive
 \begin{equation*}
 \mu\Big(\bigcup_i A_i\Big) = \sum_i \mu(A_i),
 \end{equation*}
 for each infinite sequence $\{A_i\}$ of disjoint measurable sets.
\end{Definition}
The triplet $(X, \mathcal{A}, \mu)$ of a set, a $\sigma$-algebra and a measure is often called a \emph{measure space}, and only $(X, \mathcal{A})$ is a \emph{measurable space}.

One of the most important and useful achievements of measure theory is the \emph{Lebesgue-integral}. For this, let us first define the notion of a measurable function, of which several similar and equivalent definitions exist, which are the functions we can actually define an integral for.
\begin{Definition}
 For $(X, \mathcal{A})$ a measurable space and $A\in\mathcal{A}$, a measurable function with respect to $\mathcal{A}$ $f$ is a function
 \begin{equation*}
 f \colon \ A \rightarrow \mathbb{R},
 \end{equation*}
 such that for every real number $t\in\mathbb{R}$, the set $\{ x \in A \mid f(x) < t\}\in \mathcal{A}$.
\end{Definition}
Two important notions used for the Lebesgue integral, and used in Section~\ref{sec:associative_constr}, are indicator functions and simple functions. An \emph{indicator function} $\mathbf {1}_A \colon X \rightarrow \mathbb{R}$ is a function that has the value $1$ on the measurable set $A$, and $0$ elsewhere, and may be seen as a generalisation of a step function. We define the integral of an indicator function as the volume of the set $A$, i.e.,
\begin{equation*}
 \int_X \mathbf{1}_A \,{\rm d}\mu(x) := \mu(A),
\end{equation*}
which makes sense since $A$ is measurable. A \emph{simple function} $f \colon X \rightarrow \mathbb{R}$ consists of a finite sum of these indicator functions for disjoints measurable sets $A_i$ and real numbers $\alpha_i$
\begin{equation*}
 f(x) = \sum_{i=1}^n \alpha_i \mathbf{1}_{A_i}.
\end{equation*}
Note that since $A_i$ are all measurable and disjoint, we can readily define an integral over these simple functions
\begin{equation*}
 \int_X f(x) \,{\rm d}\mu(x) := \sum_{i=1}^n \alpha_i \mu(A_i).
\end{equation*}
Let us denote the collection of all simple functions on $(X, \mathcal{A})$ as $\mathcal{S}$. The Lebesgue-integral for a~positive valued function $f_+(x) \geq 0$ is now defined as
\begin{equation*}
 \int_X f_+(x) \,{\rm d}\mu(x) := \sup\left\{ \int_X g(x) \,{\rm d} \mu(x)\mid g \in \mathcal{S},\, g(x) < f_+(x) \, \forall x\in X\right\}.
\end{equation*}
This can be seen as approximating a function $f_+$ by simple functions. For a general measurable function $f$ we then split the function up into two positive functions $f = f_+ - f_-$, $f_+ \geq 0$, $f_- \geq 0$. Finally, the Lebesgue integral for any measurable function is defined as
\begin{equation*}
 \int_X f(x) \,{\rm d}\mu(x) := \int_X f_+(x) \,{\rm d}\mu(x) - \int_X f_-(x) \,{\rm d}\mu(x).
\end{equation*}

The Lebesgue integral is a more general notion of integration than the Riemann integral, but importantly, if the Riemann integral of a function $f(x)$ exists, the Lebesgue integral also exists and they are equal.

\section{The Laplace--Beltrami operator}\label{sec:app:laplace}
This section serves as an introduction to the Laplace--Beltrami operator, which generalises the Laplacian from Euclidean space $\mathbb{R}^n$ to Riemannian manifolds $(\mathcal{M}, g)$. More information on this operator may be found in literature, for instance~\cite{rosenberg_1997}. Since we are mainly interested in smooth functions in this work, we will always consider the function spaces of smooth functions $C^\infty(\mathbb{R}^n)$ and $C^\infty(\mathcal{M})$, respectively.

The Laplacian on Euclidean space $\mathbb{R}^n$ is an operator
\begin{equation*}
 \Delta \colon \ C^\infty(\mathbb{R}^n) \rightarrow C^\infty(\mathbb{R}^n),
\end{equation*}
and is given by
\begin{equation*}
 \Delta := \partial_{1}^2 + \dots + \partial_{n}^2,
\end{equation*}
and it shows up in various contexts throughout physics. If one wishes to generalise this operator to a Riemannian manifold $(\mathcal{M}, g)$, one needs to define it in a coordinate-free way. This is achieved as
\begin{equation*}
 \Delta := {\rm div} \circ \nabla.
\end{equation*}
Here $\nabla$ denotes the gradient of a function, and ${\rm div}$ the divergence. This definition now works for general Riemannian manifolds. Locally, it is given by
\begin{equation*}
 \Delta f = \frac{1}{\sqrt{\det g}} \partial_j \big(g^{ij} \sqrt{\det g} \partial_i f\big).
\end{equation*}

On an $n$-dimensional Riemannian manifold, one can define a measure by the integration form
\begin{equation*}
 {\rm d}^n x \sqrt{\det g}.
\end{equation*}
Using this integration form, one can define an inner product on $C^\infty(\mathcal{M})$,
\begin{equation*}
 \braket{f| g} := \int_\mathcal{M} {\rm d}^n x \sqrt{\det g} f(x) g(x).
\end{equation*}
Taking the closure of $C^\infty(\mathcal{M})$ with respect to this inner product leads to the Hilbert space of square integrable functions $L^2(\mathcal{M})$. Using this inner product, the Laplace--Beltrami operator is actually self-adjoint for compactly supported functions~\cite{STRICHARTZ198348}
\begin{equation*}
 \braket{\Delta f | g} = \braket{f| \Delta g}.
\end{equation*}

A final remark goes to the link to geometry~\cite{rosenberg_1997}. Not only is the Laplace--Beltrami operator determined by the metric, but this works the other way around too. Knowing how the Laplace--Beltrami operator acts on functions actually fixes the metric completely.

\section{The canonical tensor model}\label{sec:app:CTM}
This section gives a brief review of the canonical tensor model (CTM). The CTM is a tensor model for gravity in the canonical (Hamiltonian) formalism, and it is the main motivation to introduce the formalism of this paper. The tensors are fully symmetric tensors of rank three, denoted by $Q_{abc}$. In the case of an $N$-dimensional underlying vector space, this means that the configuration space is $\mathbb{R}^\mathcal{N}$, where $\mathcal{N}=\frac{1}{6}N(N+1)(N+2)$ is the amount of independent entries in the tensor.

Since the model is constructed in the canonical formalism, we construct a phase space which is isomorphic to $\mathbb{R}^{2\mathcal{N}}$, where the canonically conjugate pair is denoted by $(Q_{abc},P_{abc})$. On this phase space, the Poisson bracket is given by
\begin{equation*}
 \{Q_{abc}, P_{abc}\} = \sum_{\sigma} \delta_{a\sigma_d} \delta_{b\sigma_e} \delta_{c\sigma_f},
\end{equation*}
and all other brackets vanish, where $\sigma$ denote the permutations of $\{d,e,f\}$.

Similarly to the ADM-formalism~\cite{Arnowitt1959}, the Hamiltonian of the theory consists of two constraints. One analogous to the spatial diffeomorphism constraint of the ADM-formalism, the generator of~${\rm SO}(N)$ transformations,
\begin{equation*}
 \mathcal{J}_{ab} = \frac{1}{4} (Q_{acd}P_{bcd} - Q_{bcd}P_{acd} ).
\end{equation*}
The other constraint, analogous to the Hamiltonian constraint in the ADM-formalism, is given by
\begin{equation*}
 \mathcal{H}_a = \frac{1}{2} ( P_{abc} P_{bde} Q_{cde} - \lambda Q_{abb} ),
\end{equation*}
where $\lambda$ is a real constant. This Hamiltonian has been shown to be unique under some physically reasonable assumptions~\cite{Sasakura:2012fb}.

Just like the ADM-formalism, the constraints span an algebra given below, which reproduces the ADM-algebra in a formal continuum limit~\cite{Sasakura:2015pxa}
\begin{gather*}
 \big\{ \mathcal{H}\big(\xi^1\big), \mathcal{H}\big(\xi^2\big) \big\} = \mathcal{J}\big(\big[\tilde{\xi}^1, \tilde{\xi}^2\big] + 2 \lambda\, \xi^1 \wedge \xi^2\big),\\
 \{\mathcal{J}(\eta) , \mathcal{H}(\xi)\} = \mathcal{H}(\eta \xi),\\
 \big\{ \mathcal{J}\big(\eta^1\big), \mathcal{J}\big(\eta^2\big) \big\} = \mathcal{J}\big(\big[\eta^1, \eta^2\big]\big).
\end{gather*}
Here $\mathcal{H}(\xi) = \mathcal{H}_a \xi_a$, $\mathcal{J}(\eta) = \mathcal{J}_{ab} \eta_{ab}$, $\tilde{\xi}_{ab} = P_{abc}\xi_c$, $\big(\xi^1\wedge\xi^2\big)_{ab} = \xi^1_a\xi^2_b - \xi^1_b\xi^2_a$ and $[\cdot,\cdot]$ denotes the matrix commutator.

In the CTM quantisation is performed by means of canonical quantisation~\cite{Sasakura:2013wza}. The fundamental variables are now mapped to the self-adjoint operators with commutators
\begin{gather*}
 Q_{abc} \rightarrow \hat{Q}_{abc}, \qquad P_{abc} \rightarrow \hat{P}_{abc},\qquad
 \{ Q_{abc}, P_{def} \} \rightarrow -{\rm i} \big[ \hat{Q}_{abc}, \hat{P}_{def} \big].
 \end{gather*}
The constraints are now given by the operators
\begin{gather*}
 \hat{\mathcal{H}}_a = \frac{1}{2} \big(\hat{P}_{abc} \hat{P}_{bde} \hat{Q}_{cde} - \lambda \hat{Q}_{abb} + i\lambda_H \hat{P}_{abb}\big),\qquad
 \hat{\mathcal{J}}_{ab} = \frac{1}{4}\big(\hat{Q}_{acd} \hat{P}_{bcd} - \hat{Q}_{bcd} \hat{P}_{acd}\big).
 \end{gather*}
The constant $\lambda_H$ depends on the ordering of the operators in the first term of the Hamiltonian constraint. Requiring the Hamiltonian constraint to be self-adjoint yields
\begin{equation*}
 \lambda_H = \frac{1}{2}(N+2)(N+3).
\end{equation*}
One nice fact of this quantisation procedure is that the algebra remains of the same form as in the classical case. Just like the usual constraints in canonical quantum gravity~\cite{Thiemann}, physical states have to satisfy
 \begin{gather*}
 \hat{\mathcal{H}}_a \ket{\Psi} = 0,\qquad
 \hat{\mathcal{J}}_{ab} \ket{\Psi} = 0.
 \end{gather*}

As mentioned in the text, there are several reasons why this is an attractive model. One of the reasons is that this model, with clear connections to gravity~\cite{Chen:2016ate,Sasakura:2014gia, Sasakura:2015pxa}, actually has known solutions to the constraint equations~\cite{Narain:2014cya,Sasakura:2013wza}, even for general dimension $N$. This means that wave-functions can be found, and gives an opportunity to actually analyse them~\cite{Kawano:2021vqc,Lionni:2019rty, Obster:2017pdq, Obster:2017dhx, Obster:2020vfo,Sasakura:2019hql} with interesting results like the emergence of tensors with certain symmetries. Though so far there was not a direct spacetime interpretation, there were clear connections to geometry~\cite{Kawano:2018pip}.

\subsection*{Acknowledgements}
The author would like to thank N.~Sasakura for all the fruitful discussions, advice and encouragement that made this work possible. Furthermore would the author like to thank the referees who gave valuable input to improve the work.

%\bibliographystyle{sigma}
%\bibliography{bibliography}

\pdfbookmark[1]{References}{ref}
\LastPageEnding
\end{document}